\documentclass[number,preprint,review,12pt]{elsarticle}

\usepackage{amssymb}

\usepackage{booktabs} 

\usepackage{rotating}

\usepackage{amsmath}

\usepackage[dvipsnames]{xcolor}

\usepackage{xurl} 

\usepackage[colorlinks,allcolors=blue]{hyperref}

\usepackage{enumitem}

\usepackage[section]{placeins} 

\usepackage{booktabs}
\usepackage{caption}
\usepackage{geometry}
\usepackage{url}
\usepackage{color}
\usepackage{xcolor}
\usepackage[utf8]{inputenc}
\usepackage[T1]{fontenc}
\usepackage{lmodern}
\usepackage{graphicx}
\usepackage[style=base,figurename=Fig.,labelfont=bf,labelsep=period]{caption}
\usepackage{subcaption}
\usepackage{amsmath}
\usepackage{tikz}
\usetikzlibrary{shapes, arrows, positioning}
\usetikzlibrary{calc,fit}
\usepackage{newtxtext,newtxmath}
\usetikzlibrary{calc,backgrounds}
\usepackage{pgfplots}

\tikzstyle{startstop} = [
  rectangle,
  draw,
  fill=gray!30,
  rounded corners,
  text width=9em,
  text centered,
  minimum height=3em,
  font=\small
]

\tikzstyle{block} = [
  rectangle,
  draw,
  fill=gray!10,
  rounded corners,
  text width=9em,
  text centered,
  minimum height=3em,
  font=\small
]

\tikzstyle{decision} = [
  diamond,
  draw,
  fill=gray!10,
  text width=7em,
  text badly centered,
  aspect=1.5,
  font=\small
]

\tikzstyle{line} = [
  draw,
  thick,
  -latex'
]

\journal{Ocean Engineering}

\begin{document}

\begin{frontmatter}



\title{Integrating Regional Ice Charts and Copernicus Sea Ice Products for Navigation Risk in Alaskan Waters}


\author{
Grant J. Peel and Ersegun D. Gedikli
}

\affiliation{organization={Department of Ocean and Resources Engineering, University of Hawai‘i at Mānoa},
            addressline={2540 Dole Street}, 
            city={Honolulu},
            postcode={96822}, 
            state={HI},
            country={USA}}
\begin{abstract}

As climate change continues to reshape marginal ice zones in the Arctic, accurate and reliable sea ice data are critical for ensuring maritime safety. This study compares regional ice charts from the Alaska Sea Ice Program (ASIP) with satellite-derived Copernicus sea ice concentration data to evaluate spatial and temporal discrepancies in ice representation across Alaskan waters from January 2010 to March 2025. Daily ASIP polygons were aligned with Copernicus grid points in a common UTM framework, and residuals were computed to quantify systematic differences. Results show that Copernicus consistently underestimates ice concentration relative to ASIP, particularly in nearshore and marginal ice zones affected by land-spillover and mixed-pixel effects such as those observed in Cook Inlet. Empirical Orthogonal Function (EOF) analysis shows that both datasets capture the same dominant physical modes of sea ice variability, with the first mode representing the annual freeze–thaw cycle and the second reflecting marginal ice-zone dynamics. To assess operational implications, vessel Automatic Identification System (AIS) data were combined with ASIP ice charts using the IACS POLARIS Risk Index Outcome (RIO) framework. Approximately 36\% of AIS observations within ice-affected waters corresponded to negative RIO values, indicating that vessels frequently operated under elevated-risk conditions. These findings demonstrate that ASIP and Copernicus provide complementary capabilities—high-resolution, analyst-driven detail and broad-scale satellite coverage—that together enable more accurate and operationally meaningful Arctic navigation and risk assessments.

\end{abstract}



\end{frontmatter}



\section{Introduction}

Climate change continues to reshape the Arctic environment, with rising temperatures leading to significant reductions in sea ice extent and thickness \citep{Serreze2015,IPCC2023,Haine2017}. Over the past several decades, the region has experienced a notable transition toward thinner, first-year ice, resulting in a broader marginal ice zone (MIZ) during the warmer months \citep{Strong2013}. Although diminished sea ice cover and longer ice-free seasons have facilitated new economic opportunities, such as resource exploration, fishing, and commercial shipping, they also bring heightened operational hazards for maritime industries \citep{ng2018implications, Boylan2021, Peel2023, Peel2024}. Ensuring that vessels have timely and accurate information about ice conditions is therefore critical for safe Arctic navigation, risk management, and cold regions infrastructure planning \citep{yang2024review}. For this purpose, satellites provide an indispensable tool for large-scale, year-round sea ice monitoring. Passive microwave sensors (e.g., SSM/I, AMSR2, SSMIS) offer daily Arctic-wide ice concentration data at spatial resolutions on the order of 25\,km \citep{Ivanova2015, Lavergne2019, Kern2020}. While these products are highly valuable for broad-scale assessments and climatological studies, they often face limitations in nearshore and complex ice environments. For example, land-contamination effects, melt pond misclassifications, and coarse grid spacing can mask narrower coastal ice edges, polynyas, and fragmented floes \citep{Meier2015a, Wagner2021}. During the melt season, meltwater on the ice surface can exhibit a microwave signature comparable to open water, causing systematic underestimation of ice concentrations \citep{Ivanova2015, Kern2020, Sallila2019}, consistent with previous comparisons between Copernicus reanalysis products and in-situ measurements at the Norströmsgrund lighthouse \citep{turner2020comparison}. Compared with high-resolution or analyst-driven charts, passive microwave algorithms have been shown to underestimate Arctic ice coverage by up to 20\%--40\% in summer \citep{Agnew2003, Meier2019}. 

In contrast, regional ice services produce analyst-interpreted, higher-resolution ice charts that capture finer-scale features in ice concentration and extent. The Alaska Sea Ice Program (ASIP), part of the U.S. National Weather Service, compiles daily charts using a blend of synthetic aperture radar (SAR), visible/infrared imagery, and in situ observations. This integrated, expert-driven process more reliably detects low-concentration ice—particularly in marginal or nearshore regions where coarse satellite products tend to be less accurate. Early validations show that ASIP’s depiction of the ice edge aligns more closely with on-the-ground observations than purely automated satellite retrievals \citep{Pacini2025}.

In parallel, recent studies highlight that accurate, higher-resolution, and continuously updated sea ice information is fundamental to risk-informed Arctic voyage planning and operational decision-making, as modern navigation safety models—including POLARIS—depend critically on reliable ice concentration and ice-type data \citep{yang2024review}. Meeting this need, however, requires understanding the performance of existing ice information sources. Despite the demonstrated effectiveness of ASIP charts for local navigation and operational planning, systematic comparisons with global satellite-based ice concentration data, such as those provided by the Copernicus Marine Environment Monitoring Service (CMEMS) \citep{CopernicusTech2016}, remain limited, particularly along the Alaskan coast. Copernicus satellite products offer timely and broad-scale Arctic coverage derived from a suite of satellite missions but typically lack the precision needed for detailed assessments in nearshore and marginal ice zones. In contrast, ASIP charts, created through expert interpretation, deliver fine-scale spatial detail, capturing critical localized ice hazards essential for vessel safety. Understanding how these two complementary data sources converge or diverge in their reported sea ice concentrations is critical for maritime stakeholders, ice forecasters, and engineers requiring reliable operational information for safe navigation, route planning, port design, and risk assessment. Beyond navigation, accurate regional ice representations are also essential inputs for estimating wave- and ice-induced loads on offshore structures and marine energy systems \citep{li2024wave, behnen2025nonlinear, li2026ice}, because even modest changes in ice concentration and marginal ice zone extent can strongly affect fatigue damage and extreme-event statistics. Therefore, analyzing both datasets together allows us to leverage the strengths of each while addressing their respective limitations.

In this study, we present a comprehensive comparison of Copernicus sea ice concentration data and ASIP ice charts across Arctic and sub-Arctic waters surrounding Alaska for the period 2010–2025. By spatially aligning the Copernicus grid with ASIP polygonal ice charts, we quantify systematic differences in how each product represents marginal, coastal, and seasonally dynamic ice regimes. To assess the operational implications of these discrepancies, we integrate Automatic Identification System (AIS) vessel data and apply the IACS POLARIS framework \citep{fedi2018} to evaluate navigation risk under varying ice conditions. This analysis reveals both the spatial extent and practical significance of mismatches between the datasets, particularly where satellite products systematically underestimate or overestimate ice concentrations relative to analyst-based assessments. By quantifying these discrepancies and examining how they propagate into ship–ice interaction risk models, the study demonstrates the value of integrating high-resolution regional ice charts with broad-scale satellite observations. Such integration is essential for improving sea ice hazard assessment, navigation planning, and the resilience of Arctic maritime operations.

\section{Data Sources}

In this study, we use two distinct ice concentration datasets for comparative analysis: satellite-derived sea ice data from the Copernicus Climate Change Service (Copernicus C3S, 2024) and high-resolution ice charts from the ASIP \cite{ASIP2024}.

\subsection{Copernicus Dataset}

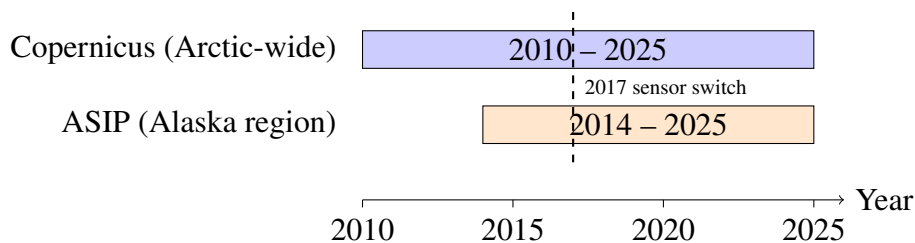
\begin{figure}[htbp!]
\centering
\begin{tikzpicture}[xscale=0.4, yscale=1]
  \draw[->] (0,0) -- (16,0) node[right]{Year};
  \foreach \x in {0,5,10,15} {
    \pgfmathtruncatemacro\year{2010 + \x}
    \draw (\x,0) -- (\x,-0.1) node[below]{\year};
  }

  \node[anchor=east] at (-0.5,2) {Copernicus (Arctic‐wide)};
  \node[anchor=east] at (-0.5,1) {ASIP (Alaska region)};

  \draw[fill=blue!20] 
    (0,2-0.25) rectangle (15,2+0.25) 
    node[midway]{2010 – 2025};

  \draw[fill=orange!20] 
    (4,1-0.25) rectangle (15,1+0.25) 
    node[midway]{2014 – 2025};

  \draw[dashed, thick] (7,2.5) -- (7,0.5) 
    node[midway, right, font=\scriptsize]{2017 sensor switch};

\end{tikzpicture}
\caption{Temporal and spatial coverage of the two main data sources.  
The top bar shows Copernicus sea‐ice products (Arctic‐wide) from January 2010 through March 2025, with a dashed line at 2017 marking the transition from AMSR‐E/AMSR2 to SSMIS sensors.  
The bottom bar shows ASIP regional ice charts around Alaska (SIGRID‐3/WMO formats), daily from June 2014 through March 2025.}
\label{fig:datasource_overlap}
\end{figure}

The Copernicus dataset consists of daily sea ice concentration products covering Arctic regions based on satellite observations. Two main data products have been utilized within this dataset, differentiated by their satellite source and time period (see Fig. \ref{fig:datasource_overlap}). Prior to 2017, sea ice data were derived from the ESA’s Climate Change Initiative (ESA CCI) medium-resolution data captured by the Advanced Microwave Scanning Radiometer Earth Observing System (AMSR-E/AMSR2) satellite missions. After 2017, the dataset transitioned to data provided by EUMETSAT through the Ocean and Sea Ice Satellite Application Facility (OSI SAF). These latter data are coarse-resolution observations collected by the Special Sensor Microwave Imager/Sounder (SSMIS) missions. In operational settings, daily updates are important to capture the trends for voyage planning and risk assessment. Both Copernicus products share the same grid resolution of \textcolor{black}{25~km}, yet the AMSR-E/AMSR2 sensors (used pre-2017) provide a finer spatial resolution and, consequently, a more detailed representation of sea ice conditions, especially within the MIZ.  In this paper, both ESA CCI and OSI SAF datasets will be collectively referred to as the ``Copernicus dataset.''

\subsection{Alaska Sea Ice Program's Ice Chart}

The ASIP dataset utilized in this study comprises ice charts provided by the National Oceanic and Atmospheric Administration’s (NOAA) NWS ASIP. The ASIP ice charts are generated daily by experienced sea ice analysts who manually delineate ice boundaries based on the previous 24-hour period, incorporating satellite imagery, SAR, optical and infrared measurements, weather forecasts, and local observations \citep{Pacini2025}. Due to this expert-driven process, the ASIP dataset can be considered a more accurate ground-truth representation of local sea ice conditions compared to satellite data \citep{Pacini2025}. ASIP ice charts have historically been produced in three distinct formats. Before October 1\textsuperscript{st}, 2015, charts were generally generated using a plain-language format. After this date, the ASIP adopted the Sea Ice Grid 3 (SIGRID-3) format, which provides more detailed and standardized ice information, enabling easier integration into comparative analyses. Mixed in throughout the range of ASIP shapefiles, including the ones available for download through the AOOS Arctic Portal follow the World Meteorological Organization's (WMO) sea ice codes. In addition to the changing formatting in the ASIP ice charts, the increased computation power allowed for analysts to create more complex shapes that better follow the ice contours. So over time the shapefiles became larger and more detailed, assumingly representing better accuracy.

The SIGRID-3 and the WMO sea ice codes are fairly similar, both using different notations of the egg code. Both contain the total ice concentration (C\textsubscript{t}), represented as a range of tenths, broken up into partial concentrations (C\textsubscript{a}, C\textsubscript{b}, C\textsubscript{c}) represented in tenths, and are sorted by ice age, with the oldest being reported first. Both contain stages of development (S\textsubscript{a}, S\textsubscript{b}, S\textsubscript{c}) and Form of ice (F\textsubscript{a}, F\textsubscript{b}, F\textsubscript{c}) that correspond with the partial concentrations. There is a well-documented conversion between the two notation sets available on the Danish Meteorological Institute (DMI) website. The Plain Language Format has a total concentration (C\textsubscript{t}) represented in the WMO code notation, but does not contain the partial concentrations like the WMO and SIGRID file structures. The Plain Language Format does contains four ice age variables that correlate to the SIGRID and WMO codes Stage of Development which can be seen in Table~\ref{tab:ice_types}. 

\begin{table}[h!]
\caption{Ice category definitions in the Plain Language Format and their corresponding SIGRID-3 and WMO egg codes}
\label{tab:ice_types}
\centering
\small
\begin{tabular}{lllll}
\toprule
\textbf{Plain Language} & \textbf{Description}           & \textbf{Thickness (cm)} & \textbf{SIGRID Code} & \textbf{WMO Code} \\
\midrule
N     & New (Nilas) ice         & < 10                 & 81, 82         & 1, 2      \\
YNG   & Young ice               & 10 to < 30           & 83             & 3         \\
FL    & Light first-year ice    & 30--70               & 87             & 7         \\
FM    & Medium first-year ice   & 70--120              & 91             & 1*        \\
FT    & Thick first-year ice    & > 120                & 93             & 4*        \\
OLD   & Old (multi-year) ice                 & —                   & 95             & 7*        \\
FAST  & Landfast ice            & —                   & —             & —        \\
OPEN  & Open water              & 0                    & 00             & 0         \\
\bottomrule
\end{tabular}
{\footnotesize\textbf{Notes:} ‘‘—’’ indicates not applicable. Asterisk (*) denotes WMO codes that do not map directly to a fixed thickness range.}
\end{table}

In Table~\ref{tab:ice_types}, there are a few entries that should be noted. The FAST and OPEN entries are more to do with the ice concentration value than the ice thickness, with 100~\% and 0~\%  ice concentration, respectively. The OLD designation, which has corresponding codes in both SIGRID and WMO notations, has no ice thickness range estimate. That is due to the difficulties in ascribing thickness values for multiyear ice that can accurately encompass both the ice growth and ice melting regimes.

\subsection{Data Gaps \& Temporal Coverage}

This study's analysis period of January 1\textsuperscript{st}, 2010 to March 31\textsuperscript{st}, 2025 contains some data gaps. The ASIP began publishing ice charts daily after June 30\textsuperscript{th}, 2014 but before then, the ASIP only published ice charts three days a week on Mondays, Wednesdays, and Fridays \citep{Pacini2025}. Excluding the non-daily portions of the ASIP dataset there are still eight data gaps: 12/23/23, 01/01/22, 12/25/21, 09/01/21, 06/06/21, 06/15/18, 06/27/17 and 04/08/16. There were also eight data gaps in the Copernicus dataset: 09/12/24 through 09/17/24, 11/09/22 and 02/20/21. These missing observations were noted, and corresponding days were excluded from the analysis to maintain clarity.

\section{Methodology}

\subsection{Study Design Overview}

\begin{figure}[htbp!]
\centering
\begin{tikzpicture}[font=\small]

\def\Router{3.0}
\def\Rinner{2.0}

\definecolor{myteal}{HTML}{66C2A5}
\definecolor{myorange}{HTML}{FDAE61}
\definecolor{myyellow}{HTML}{FFD92F}
\definecolor{mygreen}{HTML}{A6D854}
\definecolor{myblue}{HTML}{74C9E3}
\definecolor{mypink}{HTML}{F9A7B0}

\begin{scope}
  \fill[myteal!80]
    (90:\Rinner) arc[start angle=90, end angle=30, radius=\Rinner] 
    -- (30:\Router) arc[start angle=30, end angle=90, radius=\Router]
    -- cycle;
  \node at ($(60:({(\Rinner+\Router)/2})$) {\parbox{2.5cm}{\centering \textbf{Start}\\[3pt]Method Setup}};
\end{scope}

\begin{scope}
  \fill[myorange!80]
    (30:\Rinner) arc[start angle=30, end angle=-30, radius=\Rinner]
    -- (-30:\Router) arc[start angle=-30, end angle=30, radius=\Router]
    -- cycle;
  \node at ($(0:({(\Rinner+\Router)/2})$) {\parbox{2.5cm}{\centering \textbf{Data Acquisition}\\(Copernicus + ASIP)}};
\end{scope}

\begin{scope}
  \fill[myyellow!80]
    (-30:\Rinner) arc[start angle=-30, end angle=-90, radius=\Rinner]
    -- (-90:\Router) arc[start angle=-90, end angle=-30, radius=\Router]
    -- cycle;
  \node at ($(-60:({(\Rinner+\Router)/2})$) {\parbox{2.5cm}{\centering \textbf{Spatial Alignment}\\(UTM \& Polygons)}};
\end{scope}

\begin{scope}
  \fill[mygreen!80]
    (-90:\Rinner) arc[start angle=-90, end angle=-150, radius=\Rinner]
    -- (-150:\Router) arc[start angle=-150, end angle=-90, radius=\Router]
    -- cycle;
  \node at ($(-120:({(\Rinner+\Router)/2})$) {\parbox{2.5cm}{\centering \textbf{Filtering}\\\& \textbf{Pre-processing}}};
\end{scope}

\begin{scope}
  \fill[myblue!80]
    (-150:\Rinner) arc[start angle=-150, end angle=-210, radius=\Rinner]
    -- (-210:\Router) arc[start angle=-210, end angle=-150, radius=\Router]
    -- cycle;
  \node at ($(-180:({(\Rinner+\Router)/2})$) {\parbox{2.5cm}{\centering \textbf{Comparative}\\\textbf{Analysis}}};
\end{scope}

\begin{scope}
  \fill[mypink!80]
    (-210:\Rinner) arc[start angle=-210, end angle=-270, radius=\Rinner]
    -- (-270:\Router) arc[start angle=-270, end angle=-210, radius=\Router]
    -- cycle;
  \node at ($(-240:({(\Rinner+\Router)/2})$) {\parbox{2.5cm}{\centering \textbf{Risk Evaluation}\\(POLARIS + AIS)\\End}};
\end{scope}

\draw[-latex, ultra thick, gray!70]
    ([shift={(95:3.2)}]0,0) arc[start angle=95, end angle=-275, radius=3.2];

\node[font=\large, align=center] at (0,0) {\textbf{Methodology}\\\textbf{Cycle}};

\end{tikzpicture}
\caption{Flowchart of the  methodology illustrating major steps. A large arrow indicates the iterative or cyclic nature of the process.}
\label{fig:creative_cycle}
\end{figure}
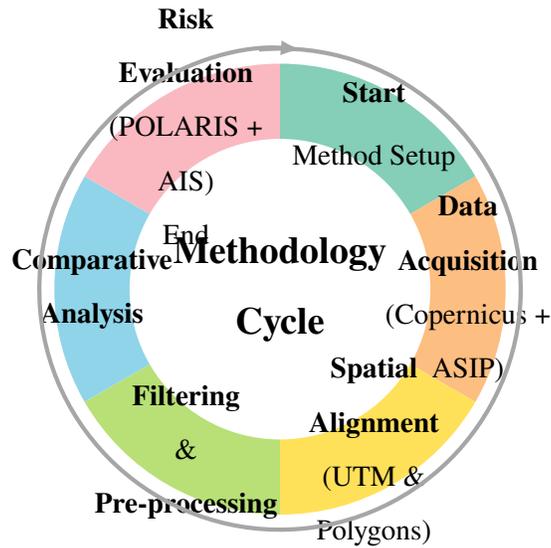

Due to inherent structural differences between the Copernicus and ASIP datasets (i.e., fixed grids for Copernicus versus dynamically changing polygon shapes for ASIP), a direct spatial comparison is challenging. To address this challenge effectively, we developed a systematic comparative approach involving spatial alignment, data filtering, and statistical analysis. Figure~\ref{fig:creative_cycle} provides an overview of the methodological cycle employed in this study, comprising of six major steps: (i) data acquisition from Copernicus and ASIP, (ii) spatial alignment in UTM zones, (iii) filtering and pre-processing, (iv) the comparative analysis itself, and (v) the final risk evaluation.

This study draws upon two primary data sources as introduced in the previous section. Table \ref{tab:comparison} illustrates the important operational details related to to both products.

\begin{table}[ht]
\centering
\caption{Overview of Data Acquisition}
\label{tab:data_acquisition}
\begin{tabular}{p{3.5cm} p{5.5cm} p{5.5cm}}
\toprule
\textbf{Feature} & \textbf{Copernicus Sea Ice Concentration} & \textbf{ASIP Ice Charts} \\
\midrule
Time Range
& Jan.~1, 2010 – Mar.~31, 2025
& Daily Coverage Jun.~30, 2014 – Mar.~31, 2025 (partial coverage Jan.~1, 2010 - Jun.~30, 2014) \\
\midrule
Sensor / Algorithm
& Pre-2017: AMSR-E/AMSR2; Post-2017: SSMIS; 25 km polar stereographic grid
& Analyst-interpreted daily shapefiles; Pre-2015: Plain-language; Post-2015: WMO SIGRID-3 \\
\midrule
Coverage \& Format
& Arctic-wide daily fields; continuous 0\%–100\%
& Alaska regional daily polygons; reported in tenths (e.g., 3–5 tenths = 30\%–50\%) \\
\midrule
Key Strengths
& Synoptic, large-area coverage; over four decades of continuity
& Fine-scale detail in nearshore/marginal zones; expert quality control \\
\midrule
Known Limitations
& Coarse spatial resolution; land-spillover and mixed-pixel effects; melt-season misclassification
& Limited spatial extent (Alaska only); occasional data gaps; small polygon overlaps \\

\bottomrule
\end{tabular}
\label{tab:comparison}
\end{table}

\subsection{Spatial Matching}

\begin{figure}[htbp!]
\centering
\includegraphics[width=1\linewidth]{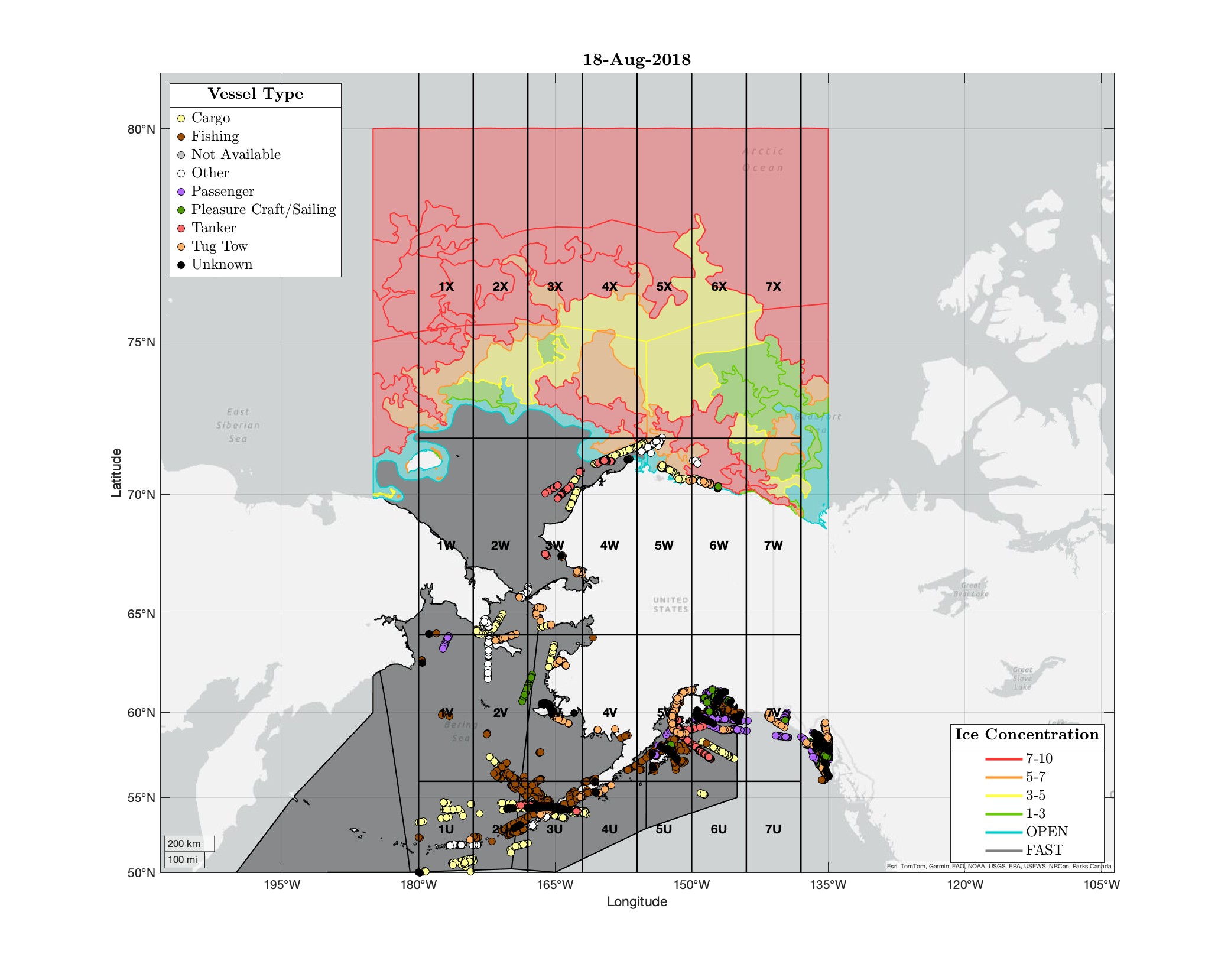}
\caption{ASIP ice chart with UTM zones and AIS data overlaid for August 18\textsuperscript{th}, 2018. Vessel types (filled circles) include Cargo (yellow), Fishing (brown), Not Available (gray), Other (white), Passenger (purple), Pleasure Craft/Sailing (green), Tanker (red), Tug Tow (orange), and Unknown (black). In terms of ice concentration, red indicates 7--10 tenths (70\%--100\%), orange indicates 5--7 tenths (50\%--70\%), yellow indicates 3--5 tenths (30\%--50\%), green indicates 1--3 tenths (10\%--30\%), blue indicates 0--1 tenths (<10\%), gray indicates landfast ice (100\%), and black indicates no ice (0\%).
}
\label{fig:utm_overlay}
\end{figure}

Copernicus sea ice concentration data reside on a fixed polar stereographic grid, whereas ASIP charts comprise irregular polygons with latitudes up to approximately 80 degrees North. These polygons are also provided in different map projections (e.g., WGS-84 geographic coordinates). Directly overlaying the two datasets without a consistent projection can introduce spatial mismatches, particularly for elongated coastal domains like Alaska's. To address this, we adopted a UTM tiling scheme, dividing the study region into 28 UTM zones (see Figure \ref{fig:utm_overlay}). This provided improved spatial consistency by minimizing geographic distortion across the Alaskan region. Initially, Copernicus grid points were filtered to include only those within Alaskan maritime areas (53~\textdegree to 75~\textdegree N, 180~\textdegree to 120~\textdegree W), excluding land points and points with substantial missing data, leaving approximately 6,097 points. Subsequently, each daily ASIP polygon was reprojected  into MATLAB's default Web Mercator Projection (WGS-84) and sorted into  UTM zones, enabling systematic point-in-polygon matching. This process involved checking each Copernicus grid point against corresponding ASIP polygons using MATLAB’s \texttt{inpolygon} function, thus associating each grid point with ASIP egg code values. This approach effectively bridged the spatial resolution gap between the datasets, giving accurate comparisons of ice conditions across the study domain while maintaining computational efficiency.

\subsection{Filtering, Pre-processing, and Data Harmonization}

Prior to conducting comparative analyses, the Copernicus and ASIP datasets underwent temporal and spatial filtering, preprocessing, and harmonization steps to ensure consistency and comparability. First, temporal harmonization was performed by removing any dates for which data were incomplete or missing in either dataset. Days lacking either Copernicus grid data or ASIP polygon coverage were excluded entirely to maintain consistent daily comparisons, resulting in a small reduction (around 1\%) in total analyzed data points. Next, due to the differing formats of ice concentration values, continuous percentage values from Copernicus versus discrete concentration ranges from ASIP, we harmonized these by converting ASIP’s discrete concentration intervals (reported in tenths, e.g., 3–5 tenths representing 30–50\%) into numerical ranges. If the Copernicus value fell within the reported ASIP range at a given location and time, this was recorded as an exact agreement. Otherwise, the discrepancy was quantified as the percentage difference to the nearest ASIP range boundary. Additionally, in cases of overlapping or closely adjacent ASIP polygons, we assigned each Copernicus grid point to the dominant ASIP polygon based on area or published priority, thereby avoiding ambiguity due to minor polygon overlaps. This comprehensive filtering, pre-processing, and harmonization approach was essential for ensuring accurate and meaningful comparative assessments of sea ice concentrations across the datasets.This harmonization provided the foundation for the discrepancy metric defined in Section~\ref{subsec:comparative_analysis}.

\subsection{Regional Subdivision}
\label{subsec:region_definitions}

\begin{figure}[htbp!]
    \centering
    \includegraphics[width=0.9\linewidth]{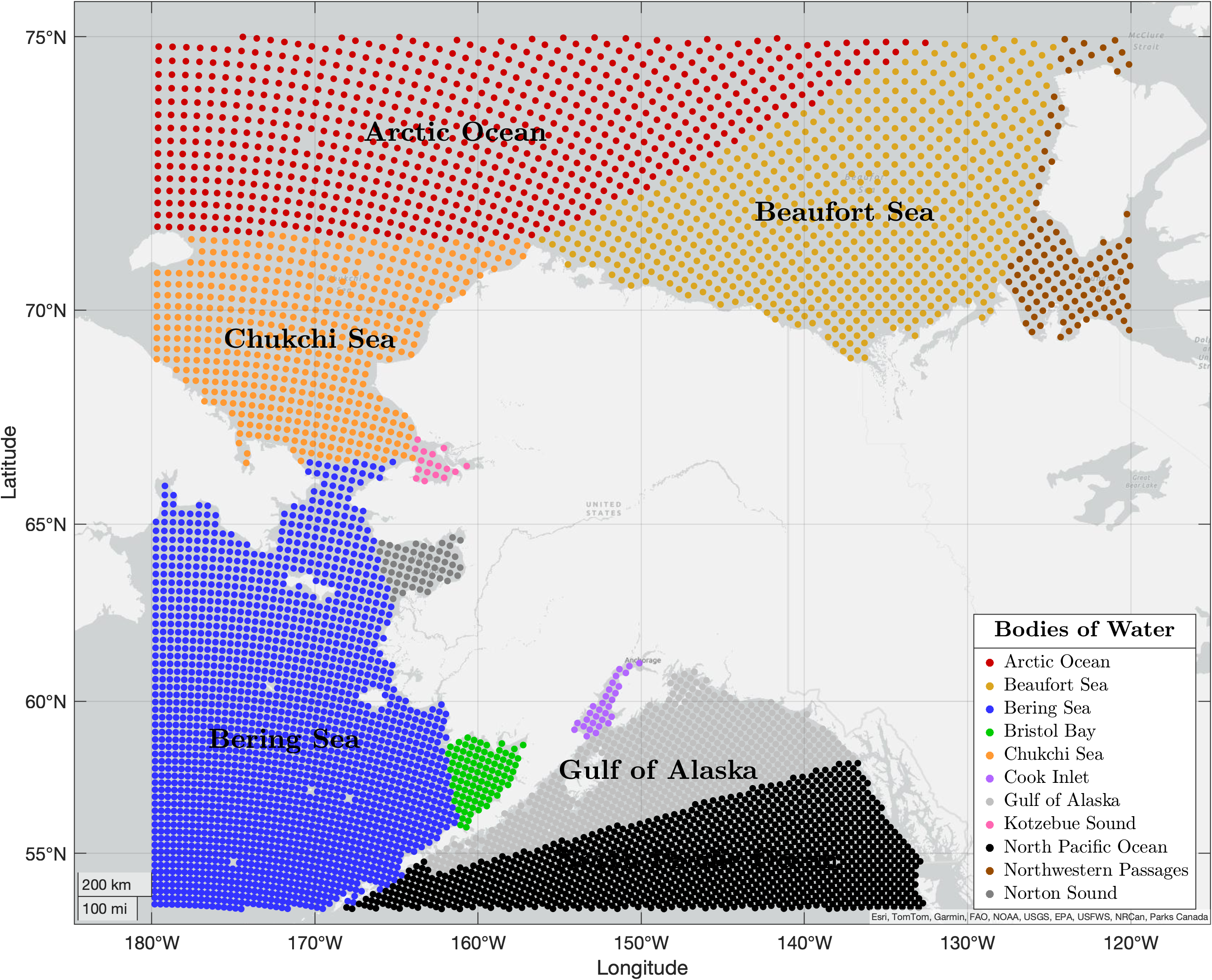}
    \caption{A visual representation of the Alaskan seas and oceans and how they are defined relative to the Copernicus grid. The colors presented for the seas in this figure are the same used in subsequent figures}
    \label{fig:Sea_Map}
\end{figure}

In order to effectively conduct a comparative analyses across Alaska's diverse maritime regions, these regions first had to be defined. The boundaries of the Arctic Ocean, Beaufort Sea, Bering Sea, Chukchi Sea, Gulf of Alaska, North Pacific Ocean, and Northwestern Passages were defined using Flanders Marine Institutes IHO Sea Areas version 3 \citep{vliz_world_seas_2021}. In addition to these oceans and seas around Alaska, we determined 4 specific regions of interest, Cook Inlet, Bristol Bay, Norton Sound and Kotzebue Sound. 

Cook Inlet is a narrow bay that connects the Port of Anchorage to the Gulf of Alaska and is bounded by the Kenai Peninsula to its east and the Alaska Range to its west. Cook Inlet is the one of the busiest Alaskan maritime regions with over half of the states population living within its watershed. While there are many boundary definitions for Cook Inlet from fisheries to natural resources, we have defined cook inlet as the water area North of the line between Cape Douglas (58.8276\textdegree~N 153.3559\textdegree~W) and Point Adams (59.2751\textdegree~N 151.9669\textdegree~W). In Figure~\ref{fig:Sea_Map}, the grid points that make up Cook Inlet are represented in purple.

Bristol Bay is a large bay on the southeastern portion of the Bering Sea bounded by the Aleutian range to the south and the Nushagak and Togiak river basins to the north. Bristol Bay is integral to Alaskan fishing fleet as it is home to the worlds largest salmon run. For this study, we defined Bristol Bay to be the water area east of the line between Nelson Lagoon (55.9962\textdegree~N 161.1946\textdegree~W) and Cape Newenham (58.7397\textdegree~N 162.1001\textdegree~W). In Figure~\ref{fig:Sea_Map}, the grid points that make up Bristol Bay are represented in green.

Norton Sound is large bay, approximately 150 miles long and 125 mile wide, on the northeast portion of the Bering Sea. Norton Sound is bounded by Seward Peninsula to the North and the Yukon river delta to the south. The Yukon River is the largest in Alaska and is navigable into Canada, making it a vital connection for many inland communities. Norton Sound's largest population center is in Nome which is located on the northern coast. For this study we defined Norton Sound to be the water area east of the line between the Yukon River Delta (62.5338\textdegree~N 165.0431\textdegree~W) and the southwester tip of Seward Peninsula (62.5338\textdegree~N 166.4383\textdegree~W). In Figure~\ref{fig:Sea_Map}, the grid points that make up Norton Sound are represented in dark grey.

Kotzebue Sound is a small arm of the Chukchi Sea, approximately 100 miles long and 70 miles wide. Kotzubue Sound is bounded by the Seward Peninsula on the south and by Baldwin Peninsula to the east. For this study we defined Kotzebue sound to be the water area southeast of the line between Cape Espenberg (66.5926\textdegree~N 164\textdegree~W) and Cape Krusenstern (67.1362\textdegree~N 163.7479\textdegree~W). In Figure~\ref{fig:Sea_Map}, the grid points that make up Kotzebue Sound are represented in pink.

\subsection{Comparative Analysis of Ice Concentration}
\label{subsec:comparative_analysis}

After harmonizing and aligning the datasets, daily comparative analyses were conducted to quantify discrepancies between Copernicus and ASIP ice concentrations. For each Copernicus grid point ($i$) and time step ($t$), we extracted the continuous concentration percentage, denoted as $C_{Copernicus}$($i,t$). This value represents the satellite-derived ice coverage estimate ranging from 0\% (completely open water) to 100\% (fully ice-covered). For the corresponding location and time, the ASIP dataset provided a concentration range due to its discrete interval format, represented as $[C_{\text{ASIP, min}},\, C_{\text{ASIP, max}}]$. These minimum and maximum concentration bounds were derived directly from ASIP polygons defined by analysts based on satellite imagery and local knowledge, and reflect uncertainty or variability within the polygonal area. A discrepancy metric, $\Delta\bigl(i,t\bigr)$, was computed for each matched point and day as follows:

\begin{equation}
\Delta\bigl(i,t\bigr)=
\begin{cases}
0, 
  & C_{\mathrm{AS},\min}\bigl(i,t\bigr)
    \;\le\;
    C_{\mathrm{Cop}}\bigl(i,t\bigr)
    \;\le\;
    C_{\mathrm{AS},\max}\bigl(i,t\bigr),\\[6pt]
C_{\mathrm{Cop}}\bigl(i,t\bigr)
  - C_{\mathrm{AS},\max}\bigl(i,t\bigr),
  & C_{\mathrm{Cop}}\bigl(i,t\bigr)
    \;>\;
    C_{\mathrm{AS},\max}\bigl(i,t\bigr),\\[6pt]
C_{\mathrm{Cop}}\bigl(i,t\bigr)
  - C_{\mathrm{AS},\min}\bigl(i,t\bigr),
  & C_{\mathrm{Cop}}\bigl(i,t\bigr)
    \;<\;
    C_{\mathrm{AS},\min}\bigl(i,t\bigr).
\end{cases}
\label{eq:delta}
\end{equation}

A value of $\Delta\bigl(i,t\bigr)=0$ indicates agreement between the datasets, meaning Copernicus ice concentration falls precisely within the range provided by ASIP. When Copernicus estimates exceed the upper bound of the ASIP range, the resulting positive discrepancy ($\Delta\bigl(i,t\bigr)>0$) represents an overestimation by the satellite-based data. Conversely, a negative discrepancy ($\Delta\bigl(i,t\bigr)<0$) occurs when Copernicus estimates are lower than the ASIP minimum, indicating underestimation. Thus, the magnitude and sign of $\Delta\bigl(i,t\bigr)$ directly quantify how satellite-derived ice concentration differs from expert-driven chart interpretations, highlighting areas and times of potential operational risk due to inaccurate ice characterization.

\subsection{Residual Calculation and Midpoint Conversion}
\label{subsec:residual_calc}

Following the formulation of the pointwise discrepancy metric in Equation~\ref{eq:delta}, additional processing was required to derive continuous residual fields suitable for temporal aggregation and statistical analysis. While the $\Delta\bigl(i,t\bigr)$ metric captures categorical agreement between Copernicus and ASIP concentration ranges, it does not directly quantify the magnitude of difference when the datasets disagree. To address this, ASIP’s range-based concentration values were converted to numerical midpoints, enabling computation of continuous residuals for each grid point and day. 

Residuals between the Copernicus and ASIP datasets were computed for each grid point and day to quantify the magnitude and sign of disagreement. Because ASIP charts report total ice concentration as discrete ranges in tenths (e.g., 3--5 tenths representing 30--50\%), a midpoint conversion was applied to facilitate direct comparison with the continuous Copernicus values. For each ASIP polygon, the midpoint value of the reported range was assigned as

\begin{equation}
C_{\mathrm{AS,mid}}\bigl(i,t\bigr) =
\frac{C_{\mathrm{AS,min}}\bigl(i,t\bigr) + C_{\mathrm{AS,max}}\bigl(i,t\bigr)}{2}.
\label{eq:midpoint}
\end{equation}

The daily residual for each grid point $i$ and time step $t$ was then calculated as

\begin{equation}
R\bigl(i,t\bigr) =
C_{\mathrm{Cop}}\bigl(i,t\bigr)
- C_{\mathrm{AS,mid}}\bigl(i,t\bigr),
\label{eq:residual}
\end{equation}

where positive residuals ($R>0$) indicate that Copernicus estimated a higher ice concentration than ASIP, and negative residuals ($R<0$) indicate underestimation. 

To maintain consistency with the earlier discrepancy metric defined in Equation~\ref{eq:delta}, the midpoint-based residuals were used for all aggregated monthly and seasonal analyses presented in Sections~\ref{subsec:seas_reg_trends}--\ref{subsec:temporal_trends}. This midpoint approach preserves the categorical nature of ASIP’s concentration ranges while enabling continuous comparison suitable for both spatial averaging and time-series aggregation.

\section{Results}

\subsection{Spatial Patterns}

\begin{figure}[htbp!]
\centering
\includegraphics[width=0.9\linewidth]{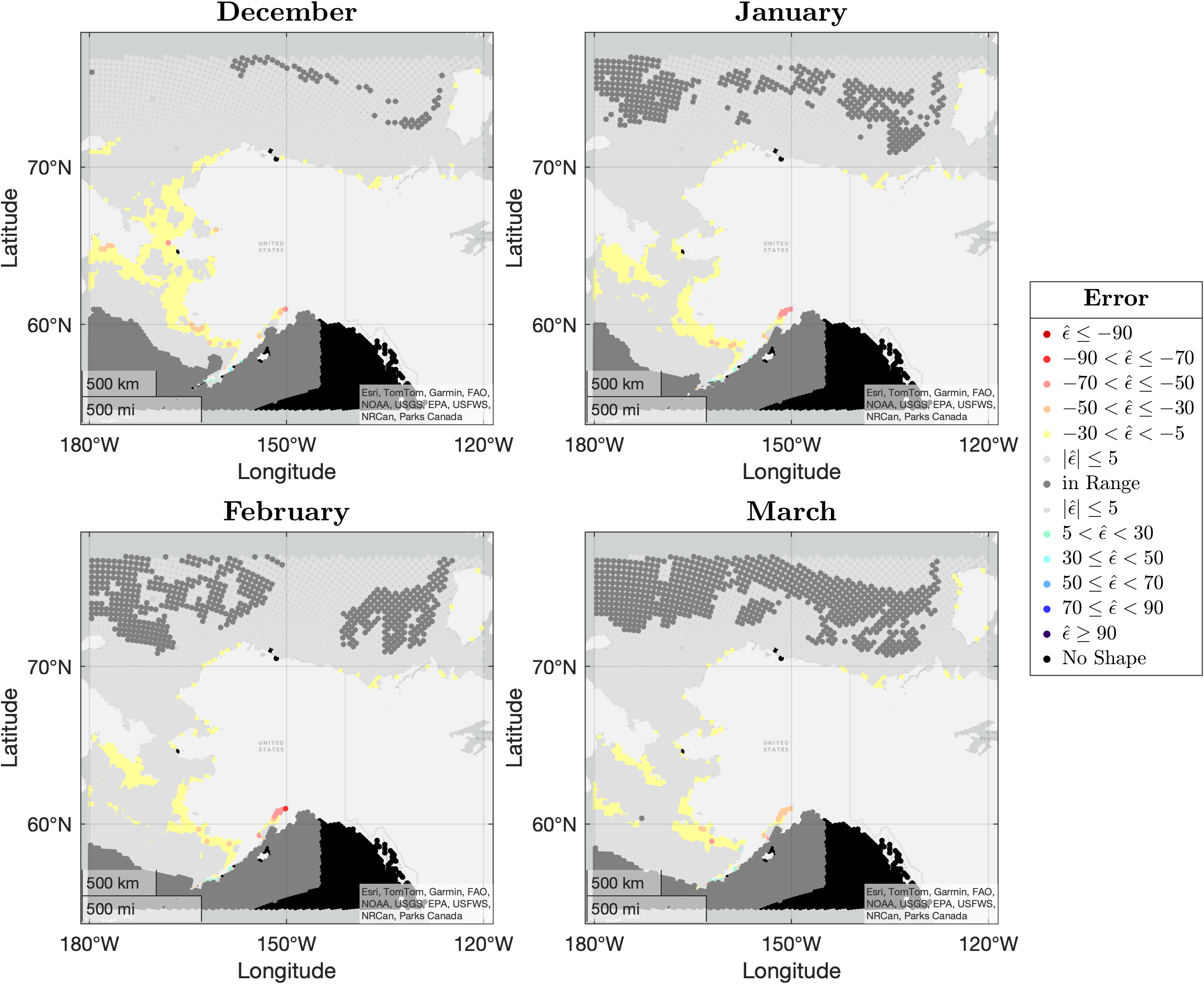}
\caption{Aggregate Monthly Average errors between the ASIP Ice Charts and Copernicus Sea Ice Concentration for the months of December (top left), January (top right), February (bottom left), and March (bottom right). The warmer colors are points when the ice charts show a higher average ice concentration than Copernicus, and the cooler colors indicate the inverse. The Black indicates points that are undefined in the ice charts, and the grey dots represent where the 2 sources agree.}
\label{fig:monthly_error_Dec_to_Mar}
\end{figure}

\begin{figure}[htbp!]
\centering
\includegraphics[width=0.9\linewidth]{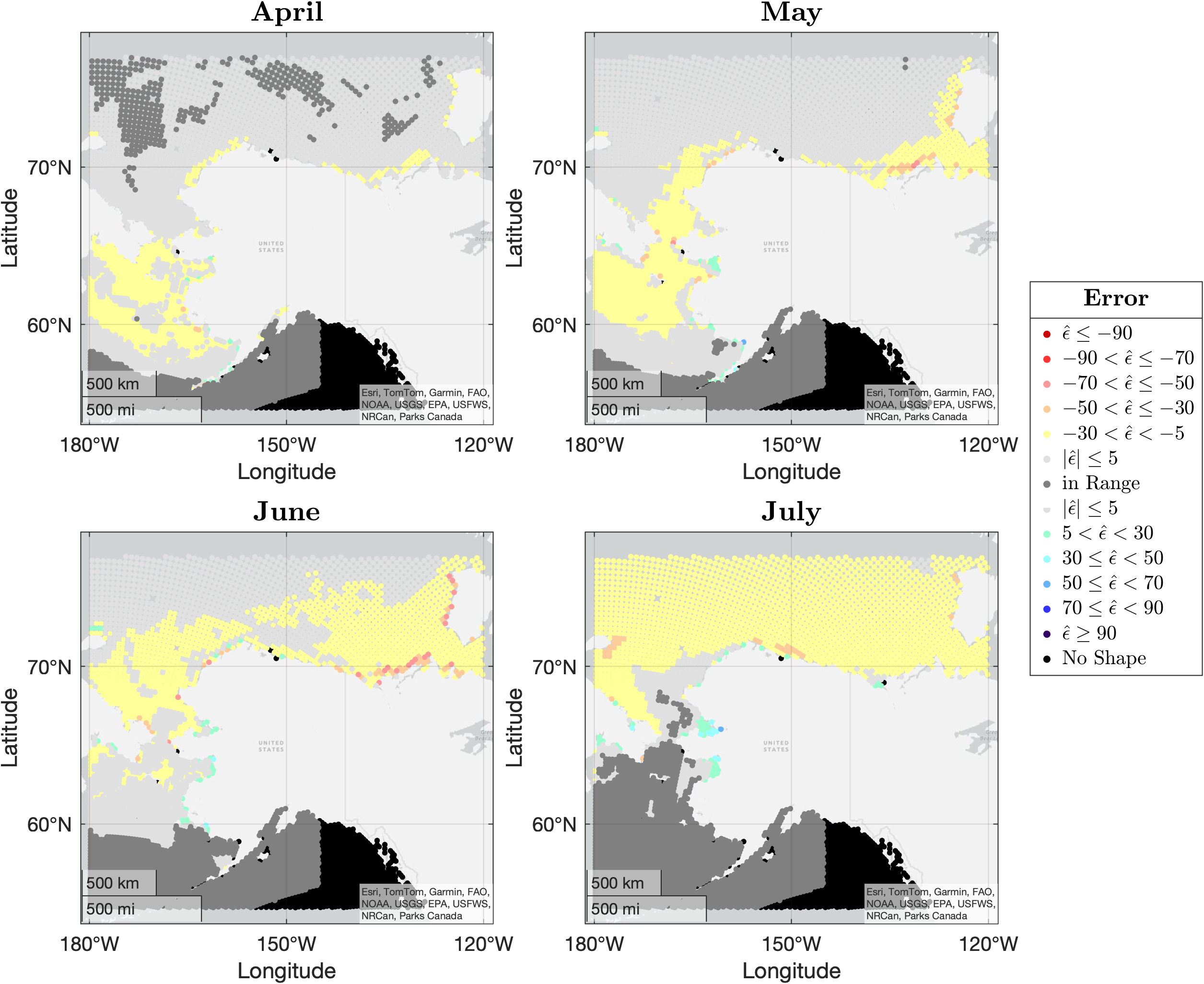}
\caption{Aggregate Monthly Average errors between the ASIP Ice Charts and Copernicus Sea Ice Concentration for the months of April (top left), May (top right), June (bottom left), and July (bottom right). The warmer colors are points when the ice charts show a higher average ice concentration than Copernicus, and the cooler colors indicate the inverse. The Black indicates points that are undefined in the ice charts, and the grey dots represent where the 2 sources agree.}
\label{fig:monthly_error_Apr_to_Jul}
\end{figure}

\begin{figure}[htbp!]
\centering
\includegraphics[width=0.9\linewidth]{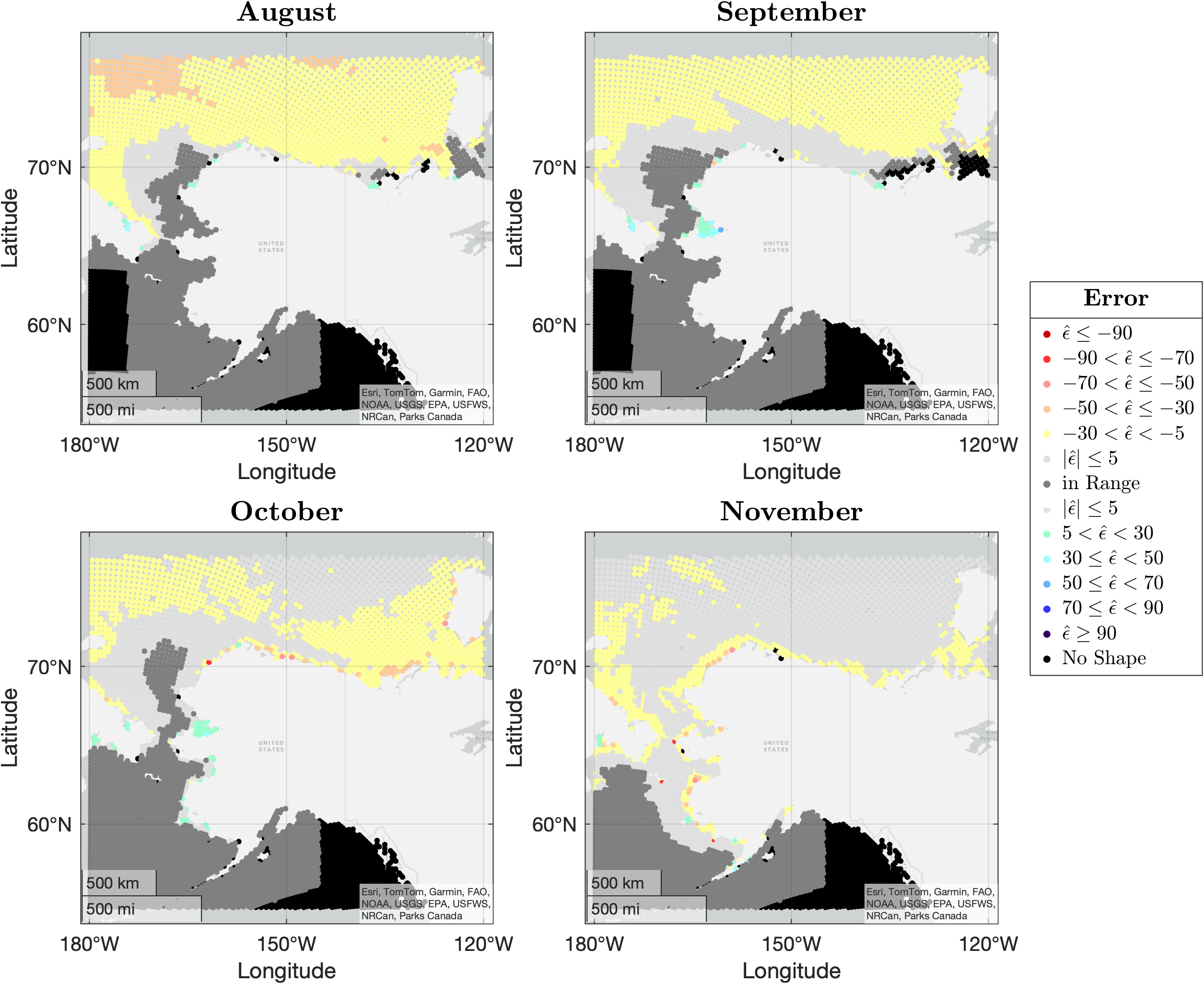}
\caption{Aggregate Monthly Average errors between the ASIP Ice Charts and Copernicus Sea Ice Concentration for the months of August (top left), September (top right), October (bottom left), and November (bottom right). The warmer colors are points when the ice charts show a higher average ice concentration than Copernicus, and the cooler colors indicate the inverse. The Black indicates points that are undefined in the ice charts, and the grey dots represent where the 2 sources agree.}
\label{fig:monthly_error_Aug_to_Nov}
\end{figure}

Following the daily comparative analysis, discrepancies between the Copernicus and ASIP ice concentration datasets were aggregated into monthly and seasonal averages to facilitate interpretation and highlight spatial patterns of error. Specifically, daily discrepancy values, $\Delta(i,t)$, at each Copernicus grid point were averaged and aggregated across each month, enabling the clear identification of persistent spatial trends and seasonal variations in ice concentration estimation (Figures~\ref{fig:monthly_error_Dec_to_Mar}--\ref{fig:monthly_error_Aug_to_Nov}). To visually illustrate these aggregated discrepancies, results were color-coded according to distinct categories. Regions consistently outside ASIP polygon coverage—representing open ocean or undefined ice areas—were shaded black, signifying no applicable ice data. Locations exhibiting exact or near-exact agreement (within $\pm$5\%) between Copernicus and ASIP datasets were shaded in shades of gray. Warm colors, ranging from yellow to red, highlighted areas where Copernicus systematically underestimated ice concentrations compared to ASIP, indicating limitations of satellite retrieval methods in accurately resolving ice conditions along ice margins, nearshore regions, and melt-affected zones. Conversely, cooler colors (blues and purples) identified systematic overestimations by Copernicus relative to ASIP, primarily occurring in regions characterized by persistent ambiguity or historical uncertainty in analyst-derived ice chart boundaries. 

These monthly maps (Figures~\ref{fig:monthly_error_Dec_to_Mar}--\ref{fig:monthly_error_Aug_to_Nov}) can be supported by statistical summaries, such as spatial averages and histograms of discrepancy distributions, offering quantitative metrics that complement and contextualize the visual findings. Here, each figure set illustrates monthly-averaged aggregated discrepancies between Copernicus sea ice concentration estimates and ASIP polygon-based charts, with individual panels representing specific months. The accompanying legend categorizes the magnitude and sign of the discrepancy ($\epsilon$) from large negative values (red hues, indicating substantial Copernicus underestimation) to large positive values (dark blues and purples, indicating substantial Copernicus overestimation). Intermediate shades of gray represent minor or negligible discrepancies, while the ``No Shape" category (black) identifies either land or areas without ASIP polygon definitions for that month.

As an illustrative example, Figure~\ref{fig:monthly_error_Dec_to_Mar} shows the discrepancy patterns from mid-winter through early spring (December to March), a period when Arctic ice coverage typically reaches its seasonal maximum. Between December and February, large contiguous areas are shaded darker gray, suggesting strong agreement between Copernicus and ASIP datasets over thick, consolidated winter ice. However, persistent red or orange patches along coastal margins, notably in Cook Inlet and nearshore regions of the Beaufort Sea, indicate ongoing underestimation by Copernicus. These discrepancies primarily stem from coarse sensor resolution, coastal geometry complexities, and land-spillover effects inherent in passive microwave retrieval methods. As conditions transition toward spring break-up in March, the appearance of more frequent yellow and red regions along the southern ice boundaries becomes evident. Nonetheless, if stable freeze conditions persist, discrepancies remain low across the central ice pack, as denoted by prevalent light and dark gray shading ($|\epsilon|\leq5\%$ and $\epsilon = 0$ respectively).

Figure~\ref{fig:monthly_error_Apr_to_Jul} illustrates the spatial distribution of monthly-averaged aggregated discrepancies between Copernicus sea ice concentration estimates and ASIP ice-chart polygons during the transitional period from late spring to mid-summer (April--July). In April and May, the maps exhibit extensive regions shaded primarily in lighter gray and yellow colors, reflecting relatively moderate discrepancies. These warmer colors indicate systematic underestimation of ice concentration by Copernicus, especially prevalent at the ice margins and coastal boundaries. For example, 1) gray and light yellow in the Bering and Chukchi Seas suggests small disagreements during the late-season ice breakup, 2) coastal areas in Cook Inlet and parts of the Beaufort Sea appear in yellow or orange, suggesting that Copernicus fails to detect or fully represent narrow or nearshore ice features. By June and July, as seasonal ice melt intensifies and the ice edge retreats northward, negative discrepancies (yellow, orange, and occasional red shading) become increasingly prominent across the Chukchi and Beaufort Seas. This intensified spatial extent of underestimation is most likely due to melt-season effects.

Figure~\ref{fig:monthly_error_Aug_to_Nov} illustrates the spatial distribution of monthly-averaged aggregated discrepancies between Copernicus sea ice concentration estimates and ASIP ice-chart polygons during the transitional period from late summer through early winter (August--November). In August and September, the Arctic reaches the minimum extent of the ice after a full melting season. The maps warmer colors in the upper latitudes generally exhibit a minor systematic underestimation of ice concentration with one exception. Kotzebue Sound flips the script showing Copernicus overestimating the ASIP. October continues this trend, implying that the error is related to the freeze up rather then the melting ice regime. Kotzebue Sound is also a narrow nearshore region with huge potential for discrepancies related to coarse sensor resolution, coastal geometry complexities, and land-spillover effects similar to Cook Inlet. By November the freezing season in the upper latitudes is in full effect and the ice estimates begin to match fairly closely again, though all of the major discrepancies occur in the coastal regions.

\subsection{Seasonal \& Regional Trends}
\label{subsec:seas_reg_trends}

\begin{figure}
    \centering
    \includegraphics[width=0.9\linewidth]{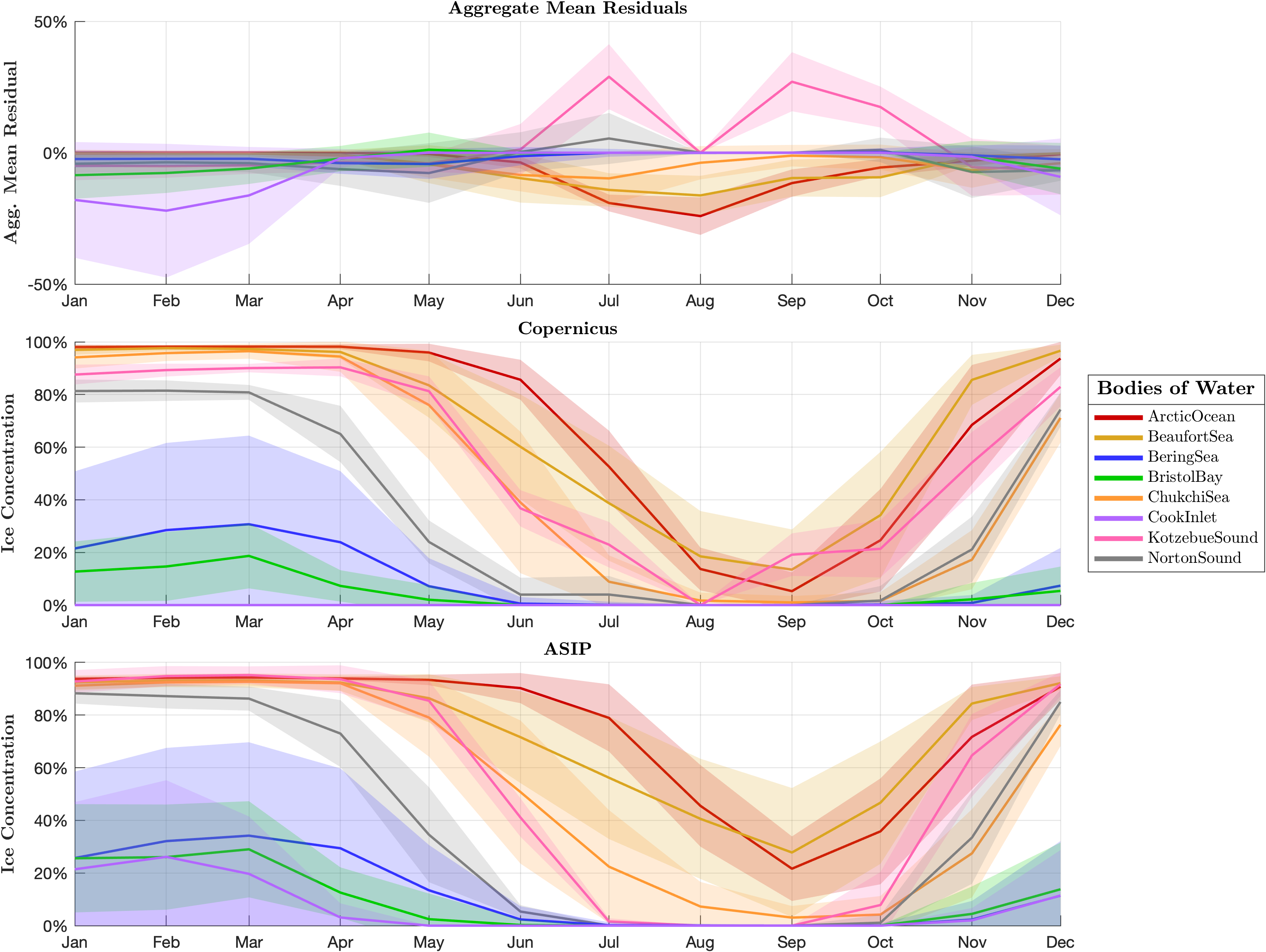}
    \caption{The top plot shows the Aggregate Monthly Mean Residuals between Copernicus Sea Ice Concentration and the ASIP Ice Charts for the different bodies of water around Alaska. The middle plot shows the average monthly Copernicus Ice Concentration for the different Alaskan bodies of water. The bottom plot shows the average monthly ASIP Ice Chart concentration for the different Alaskan bodies of water. The average for each body of water is banded by a standard deviation in each direction.}
    \label{fig:seasonal_by_region}
\end{figure}

Monthly mean discrepancies were averaged across major Alaskan seas and coastal regions to assess seasonal and regional patterns (Fig.~\ref{fig:seasonal_by_region}). Across nearly all regions, Copernicus underestimated ice concentration relative to ASIP during spring and summer (May--August) and showed close agreement during the winter maximum (January--March). The largest negative residuals occurred in Cook Inlet during winter, where irregular coastlines, mixed pixels, and tidal leads reduce the reliability of passive-microwave retrievals. Another region with consistently large negative residuals was the Arctic Ocean during summer, where melt ponds on the ice surface reduce emissivity and cause the satellite algorithm to misclassify ice as open water. The range-based concentration format of ASIP enables its analysts to represent this uncertainty more effectively than the fixed-value Copernicus grid. 

The only notable positive residuals occur in Kotzebue Sound during summer, where Copernicus detects ice that is not reported in the ASIP charts. This discrepancy likely arises from land-spillover and coarse spatial resolution, which cause nearshore pixels to appear partially ice-covered even in ice-free conditions.

\begin{figure}[htbp!]
    \centering
    \includegraphics[width=0.9\linewidth]{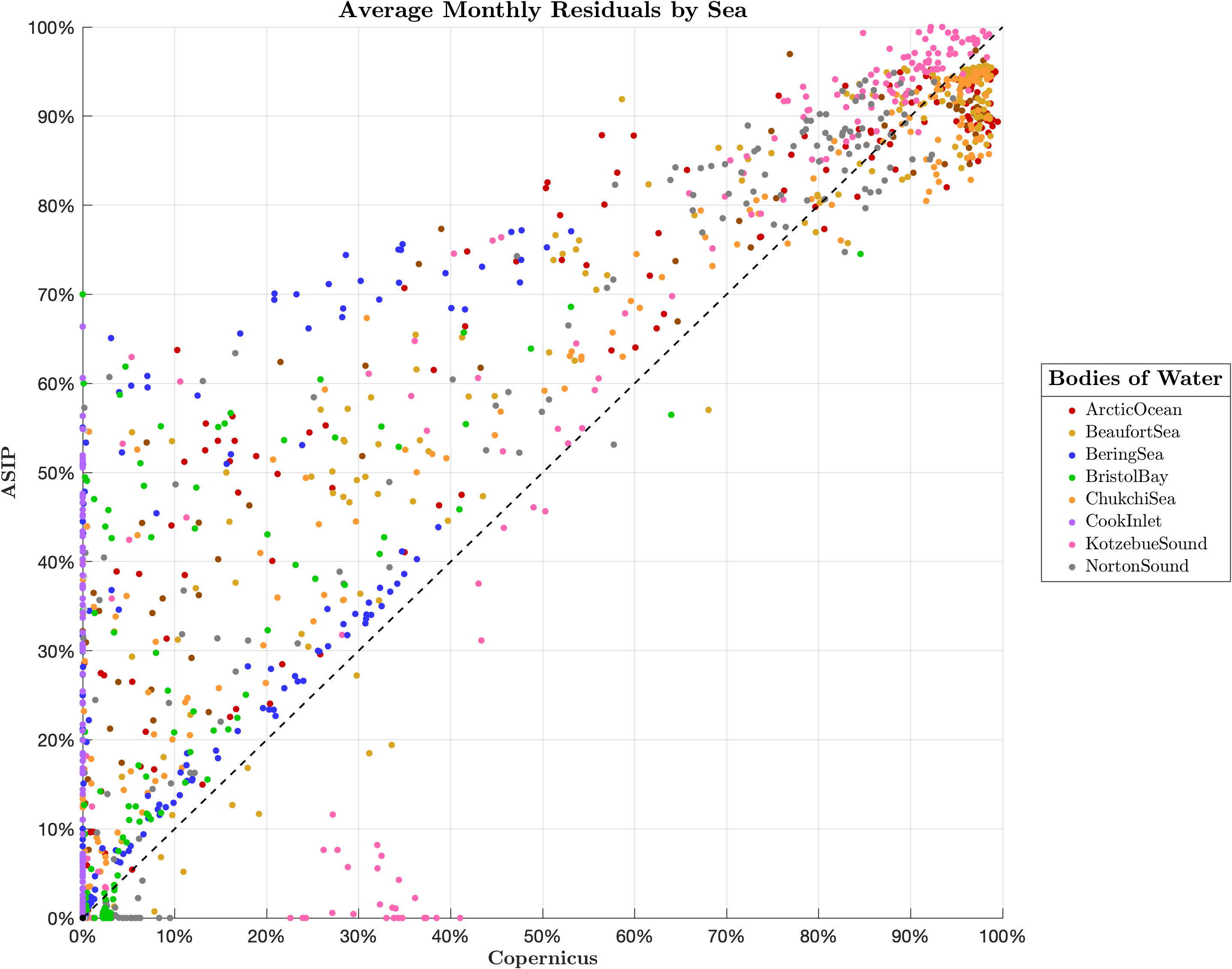}
    \caption{A direct comparison of the Average Monthly Ice Concentration for each dataset across the Alaskan bodies of water from 2010 to 2025. Each point represents the spatial average over the seas and a monthly temporal average.}
    \label{fig:scatter_monthly}
\end{figure}

Figure~\ref{fig:scatter_monthly} provides a direct comparison of monthly mean ice concentrations between the two datasets. It shows that generally, the ASIP (on the y-axis) is a more conservative estimate of the ice concentrations. Excluding the grouping of pink Kotzebue Sound points in the bottom of the figure, most of the points where Copernicus has a more conservative average ice concentration occurs in the upper right quadrant in the 80\% to 100\% range. The upper grouping on the Copernicus side of the one-to-one line mostly contains points from the Arctic Ocean, Beaufort Sea, and Chukchi Sea during times of high ice concentrations. This anomaly seems to be a product of the limitations of taking an average of a range rather than the Copernicus dataset actually being a more conservative metric. 

Taken together, these seasonal and regional results indicate that Copernicus systematically underestimates ice extent in dynamic or coastal regions, whereas ASIP provides higher and more spatially variable concentrations consistent with analyst interpretation.

\subsection{Temporal Trends}
\label{subsec:temporal_trends}

\begin{figure}[htbp!]
    \centering
    \includegraphics[width=0.9\linewidth]{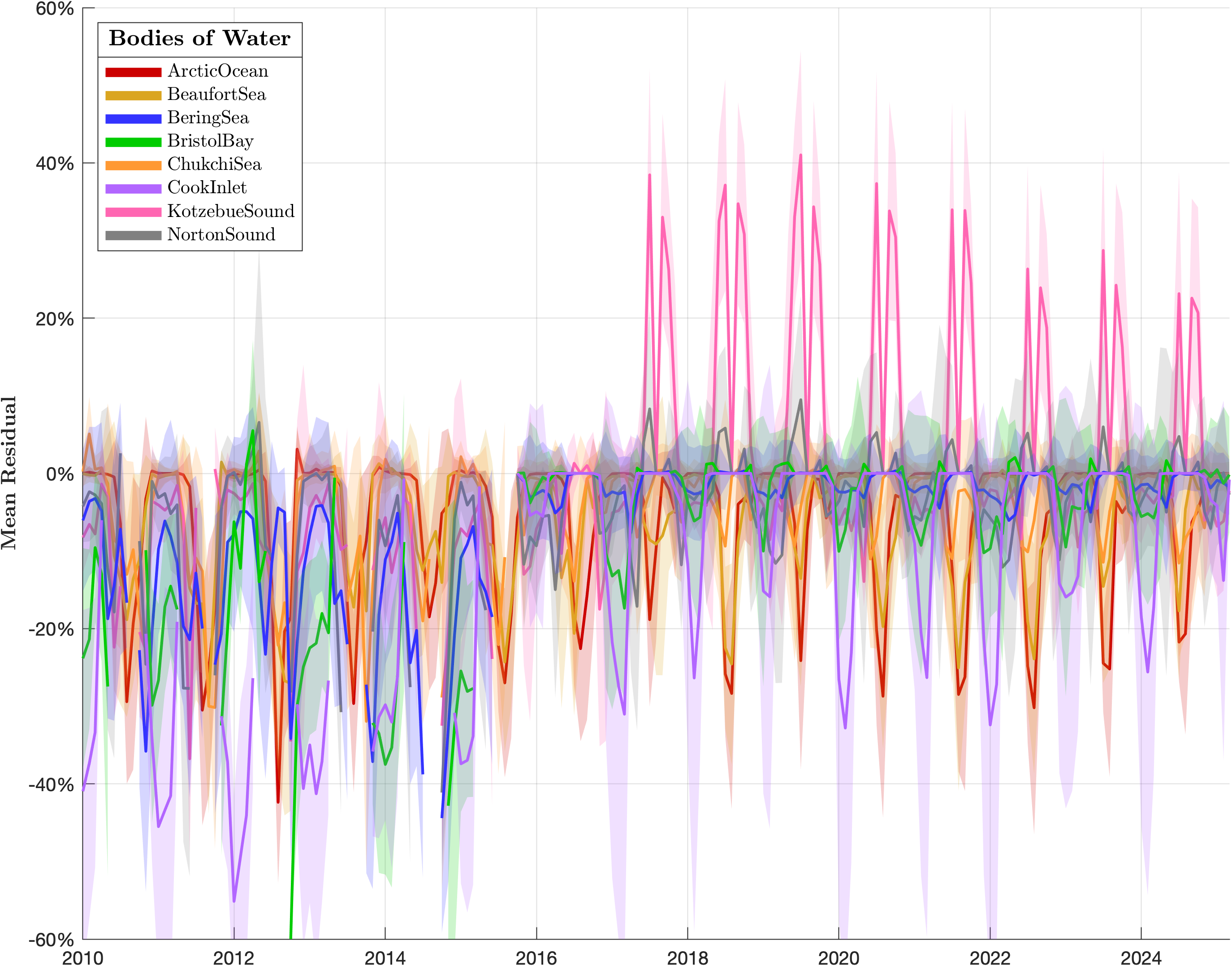}
    \caption{Monthly ice concentration residual averaged over all the grid points in the sea. The shaded regions shows one standard deviation either side of the mean.}
    \label{fig:residual_timeseries}
\end{figure}

Long-term consistency between datasets was evaluated by averaging monthly mean residuals across all grid points within each sea from 2010 to 2025 (Fig.~\ref{fig:residual_timeseries}). The resulting time series reveals a persistent negative residual---indicating Copernicus underestimation---through most of the record. After 2017, however, several regions, most notably Kotzebue Sound, display a shift toward positive residuals. This change coincides with the transition from the AMSR-E/AMSR2 sensors to the coarser-resolution SSMIS instrument in the Copernicus data stream. The reduced spatial resolution amplifies land-spillover and mixed-pixel effects, particularly in confined nearshore basins.

Inter-annual variability remains low during mid-winter, when both datasets depict stable, consolidated ice cover, but increases sharply during melt and freeze-up transitions. These fluctuations reflect the sensitivity of both observational systems to rapid seasonal changes in ice concentration and surface conditions.

Overall, discrepancies between datasets were smallest during peak winter coverage and largest during summer melt and transitional periods, particularly along the MIZ. These seasonal biases are consistent with known limitations of passive microwave retrieval algorithms and emphasize the complementary strengths of ASIP and Copernicus for different spatial and temporal scales.

\section{Discussion}

The comparative analysis revealed both seasonal and regional discrepancies between the Copernicus and ASIP datasets. Overall, Copernicus performs reliably in representing large-scale ice distribution patterns but systematically underestimates concentrations in nearshore and marginal ice zones. This underestimation is most pronounced during the melt season and within coastal regions such as Cook Inlet and the southeastern Bering Sea, where complex geometries and mixed-pixel effects reduce satellite accuracy. In contrast, ASIP’s analyst-based approach provides greater spatial detail and better alignment with known physical ice dynamics.

\subsection{Marginal Ice Zone Behavior}

Discrepancies between the datasets are strongly associated with the position and dynamics of the Marginal Ice Zone (MIZ). During the transitional months of April, May, October, and November, a band of moderate negative residuals (5--25\%) migrates seasonally across the Bering and Chukchi Seas, following the ice edge. This moving discrepancy band corresponds to the seasonal advance and retreat of the MIZ. 

During the summer months (June--August), underestimation by Copernicus becomes more widespread, with negative residuals of 5--50\% across the Western Beaufort and Eastern Chukchi Seas. These errors can be attributed to melt-pond formation, which reduces surface emissivity and causes passive microwave retrieval algorithms to misclassify partially ice-covered pixels as open water \citep{Roesel2012, cavalieri1990, comiso1996}. In contrast, winter months (January--March) exhibit narrower bands of error, indicating reduced interannual variability and greater consistency between datasets when ice coverage is consolidated. The consistent negative bias suggests that ASIP’s estimates are more conservative in representing ice coverage near the dynamic MIZ boundary.

\subsection{Regional Discrepancies}

Two regions—Cook Inlet and Kotzebue Sound—highlight contrasting manifestations of dataset disagreement. In Cook Inlet, Copernicus consistently fails to detect ice, while ASIP routinely identifies seasonal coverage. This underestimation results from three primary factors: (i) land contamination in narrow channels, (ii) smoothing of boundary gradients in the satellite retrieval algorithm, and (iii) mixed-pixel averaging along complex coastlines \citep{cavalieri1996, kern2022, spreen2008}. The result is a persistent operational gap in Copernicus data for one of Alaska’s busiest maritime regions.

Kotzebue Sound, on the other hand, shows a distinct temporal shift. Prior to 2017, discrepancies between the datasets were minimal, but after the transition to the SSMIS sensor, Copernicus began to overestimate ice coverage relative to ASIP. This inversion corresponds to the coarser spatial footprint of the SSMIS instrument, which increases susceptibility to land-spillover effects. The sensor change therefore introduced an observable bias that must be accounted for in future multi-sensor climatologies.

\subsection{Empirical Orthogonal Functions}

To examine dominant spatiotemporal modes of variability, Empirical Orthogonal Function (EOF) analysis was applied to both datasets. EOF analysis, analogous to POD and PCA based approaches used in laboratory wave-ice and ice-induced vibration studies \citep{li2019identifying,gedikli2019pressure,li2021laboratory}, decomposes sea ice concentration fields into orthogonal spatial patterns and their corresponding temporal amplitudes, allowing key variability modes to be objectively identified. 

Prior to EOF computation, both datasets were formatted and interpolated to remove missing values. Because EOF cannot handle undefined data or concentration ranges, ASIP’s discrete categories were converted to midpoint values, and temporal gaps were linearly interpolated. For summer periods and pre-SIGRID records, missing values were replaced with zeros. Points containing more than 10\% missing data were excluded, leaving 4,710 of the original 6,097 grid points. Area weighting by latitude was applied following Hannachi (2007) \cite{hannachi2007} to account for the non-uniform distribution of grid cells.

\begin{figure}[h!]
    \centering
    \includegraphics[width=0.9\linewidth]{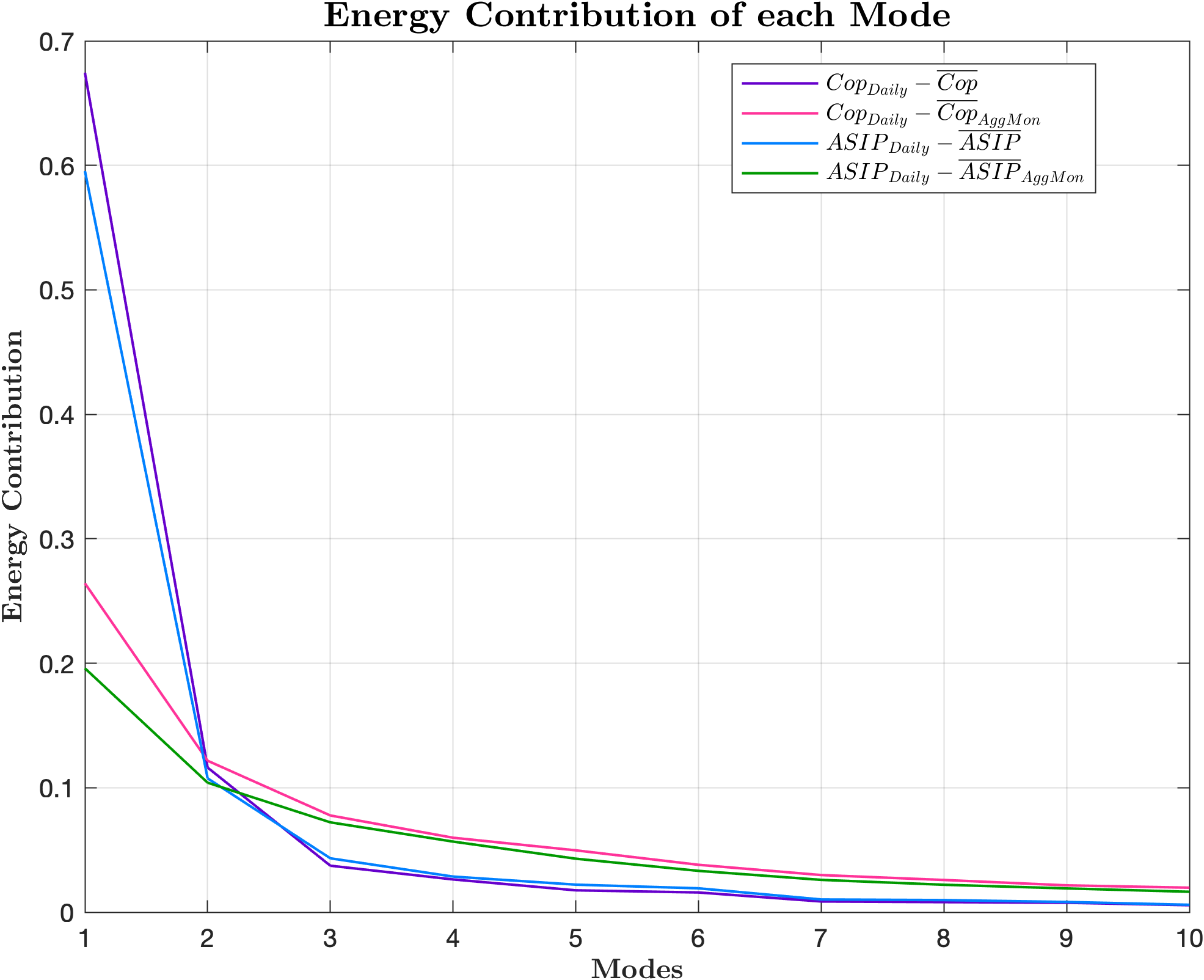}
    \caption{Relative energy contribution of each mode. The purple and pink lines represent the EOF run on the Copernicus dataset and the blue and green lines are for the ASIP. The purple and blue lines use the traditional EOF method of normalizing the dataset around the mean while the pink and green lines remove the aggregate monthly means from datasets.}
    \label{fig:eof_energy_contribution}
\end{figure}

EOF analysis was performed under two normalization schemes: (1) the traditional mean-centered method ($X_{daily} - \bar{X}$) and (2) a seasonally detrended method that removes aggregate monthly means ($X_{daily} - \bar{X}_{agg\,mon}$). The traditional EOF captured dominant seasonal variability, while the detrended EOF isolated shorter-term, non-seasonal fluctuations. Figure~\ref{fig:eof_energy_contribution} summarizes the relative variance explained by each mode.

\begin{figure}[h!]
    \centering
    \includegraphics[width=0.9\linewidth]{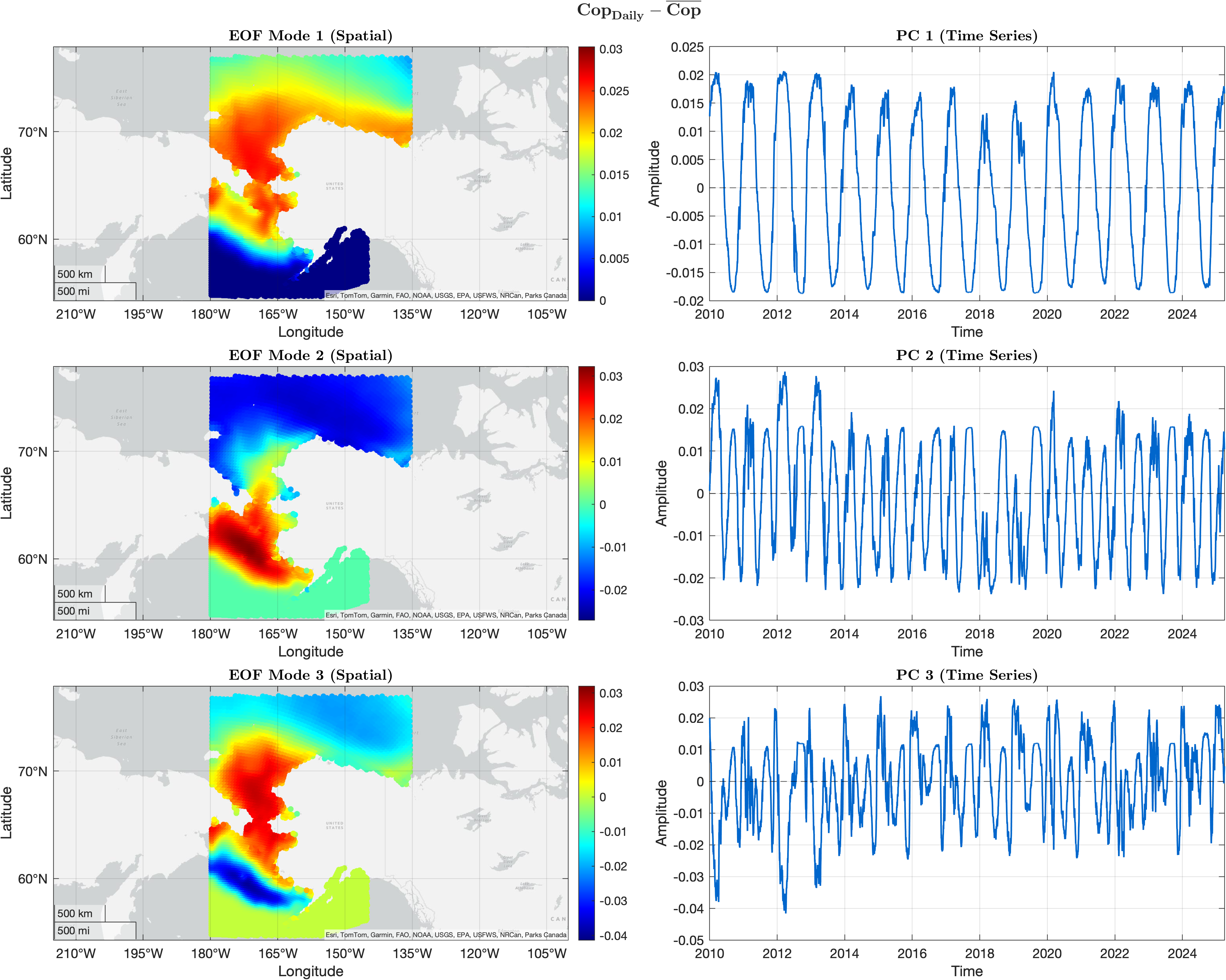}
    \caption{A temporal and spatial visualization of the first 3 modes of the Empirical Orthogonal Function performed on the Copernicus dataset normalized around its mean.}
    \label{fig:eof_copernicus}
\end{figure}

\begin{figure}[h!]
    \centering
    \includegraphics[width=0.9\linewidth]{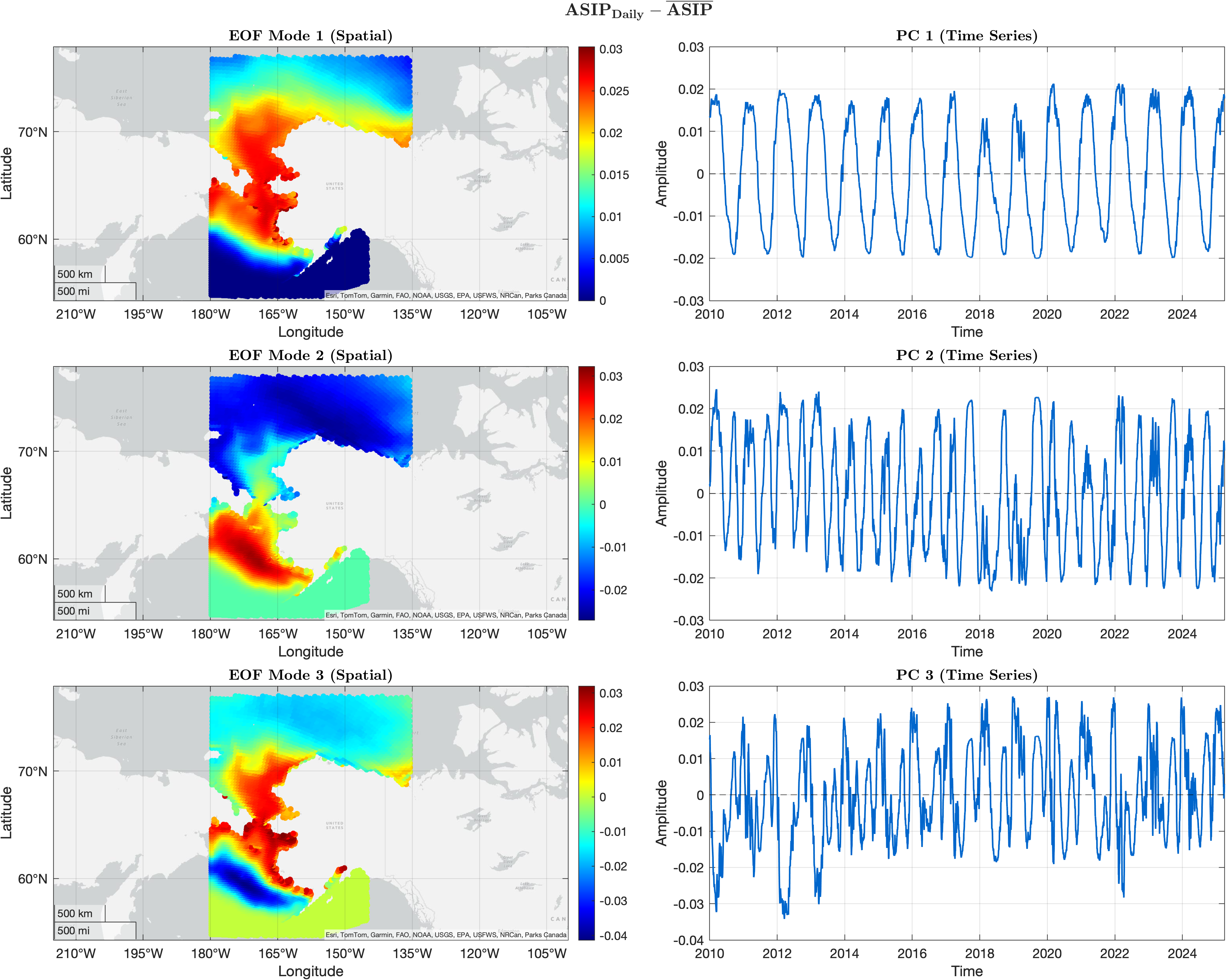}
    \caption{A temporal and spatial visualization of the first 3 modes of the Empirical Orthogonal Function performed on the Alaska Sea Ice Program dataset normalized around its mean.}
    \label{fig:eof_asip}
\end{figure}

Figures~\ref{fig:eof_copernicus} and \ref{fig:eof_asip} show the first three spatial and temporal modes derived from the traditional EOF for the Copernicus and ASIP datasets, respectively. The similarity between the corresponding modes indicates that both datasets capture nearly identical large-scale variability patterns. This consistency also holds for the seasonally detrended method (not shown), further confirming that both data sources represent the same dominant physical processes despite differences in spatial resolution and data derivation. Figure~\ref{fig:eof_energy_contribution} summarizes the relative variance explained by each mode.

In the traditional approach, the first mode accounts for approximately 60--70\% of total variance, primarily reflecting the annual seasonal cycle. This is evident from the mode’s time series, which displays a single peak and trough per year. The spatial distribution of this mode exhibits low variability in regions of perennial open water or continuous ice, and high variability along the marginal ice zone where the ice edge advances and retreats annually. When seasonal variability is removed, the first mode of the detrended EOF explains only 20--25\% of the total variance and closely matches the spatial pattern of the second mode in the traditional EOF. This correspondence confirms that the leading EOF mode in the unadjusted analysis predominantly represents the seasonal cycle rather than interannual variability.

The strong similarity of EOF modes across datasets indicates that both Copernicus and ASIP capture the same dominant physical processes driving Arctic sea ice variability, despite their structural and methodological differences.

\subsection{POLARIS Based Operational Risk Evaluation}
\label{subsec:polaris}

The EOF analysis demonstrates that both datasets consistently represent the large-scale physical dynamics of Arctic sea ice, but it does not directly address the operational implications of their differences. To translate these statistical findings into practical context, we next examine how variations in ice concentration between datasets influence real-world navigation risk through the POLARIS framework.

To assess the real-world implications of dataset discrepancies, vessel movements derived from AIS data were overlaid with ASIP-based ice concentration fields. Navigational risk was quantified using the IACS Polar Operational Limit Assessment Risk Indexing System (POLARIS), which expresses a Risk Index Outcome (RIO) for vessel–ice interactions \citep{fedi2018}. Positive RIO values correspond to normal operations, while negative values indicate elevated operational risk. The IACS recommends maximum operating speeds of 3~kn for non–ice-strengthened (NIS) vessels and 5~kn for Polar Class 3–5 vessels during such conditions.

\begin{figure}[h!]
\centering
\includegraphics[width=0.9\linewidth]{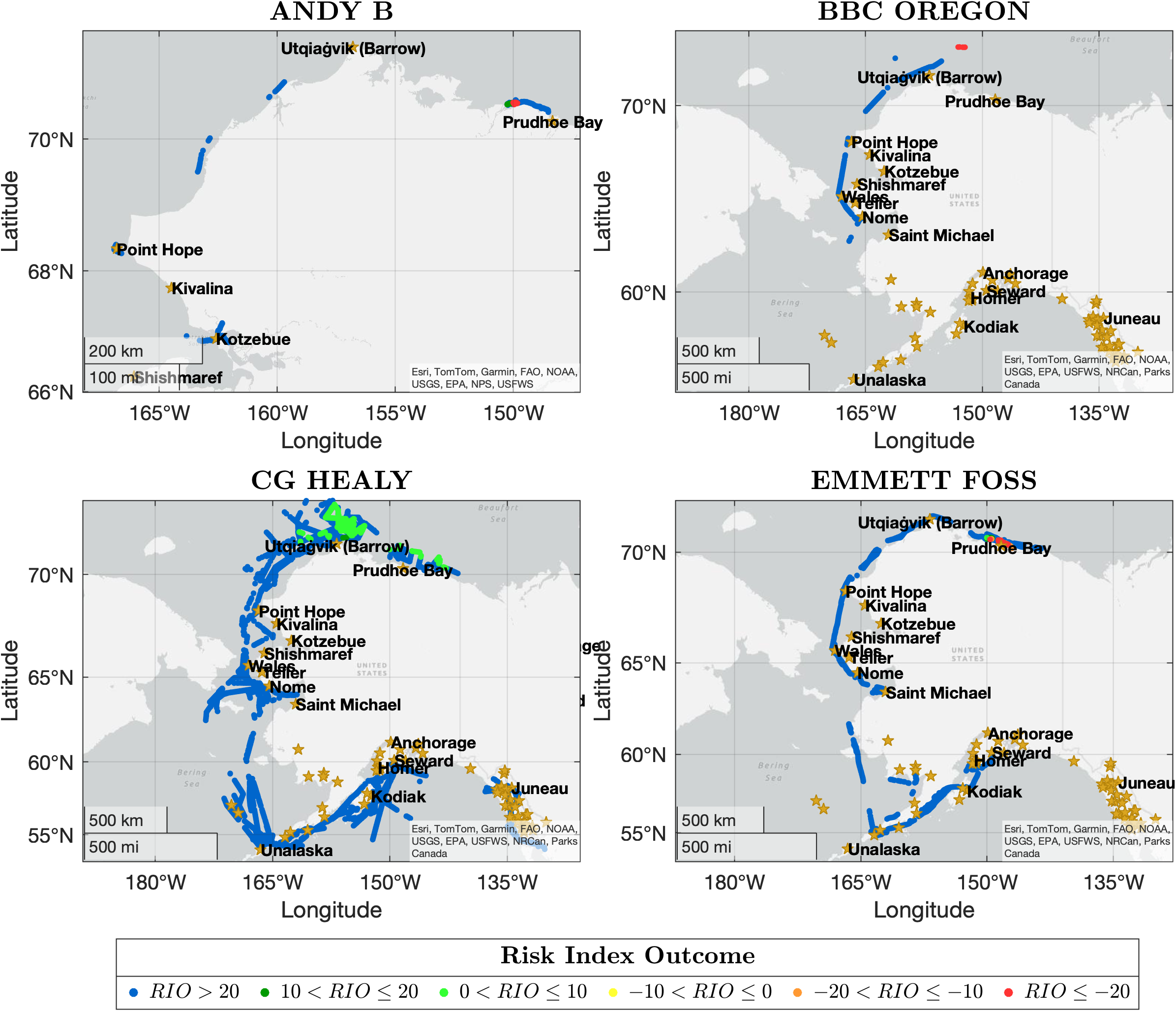}
\caption{Spatial distribution of Risk Index Outcomes (RIO) for the ships Andy B (top left), BBC Oregon (top right), CG Healy (bottom left) and the Emmett Foss (bottom right). Negative RIO values indicate elevated operational risk. Gold stars represent public Alaskan port facilities.}
\label{fig:rio_map_1}
\end{figure}

\begin{figure}[h!]
\centering
\includegraphics[width=0.9\linewidth]{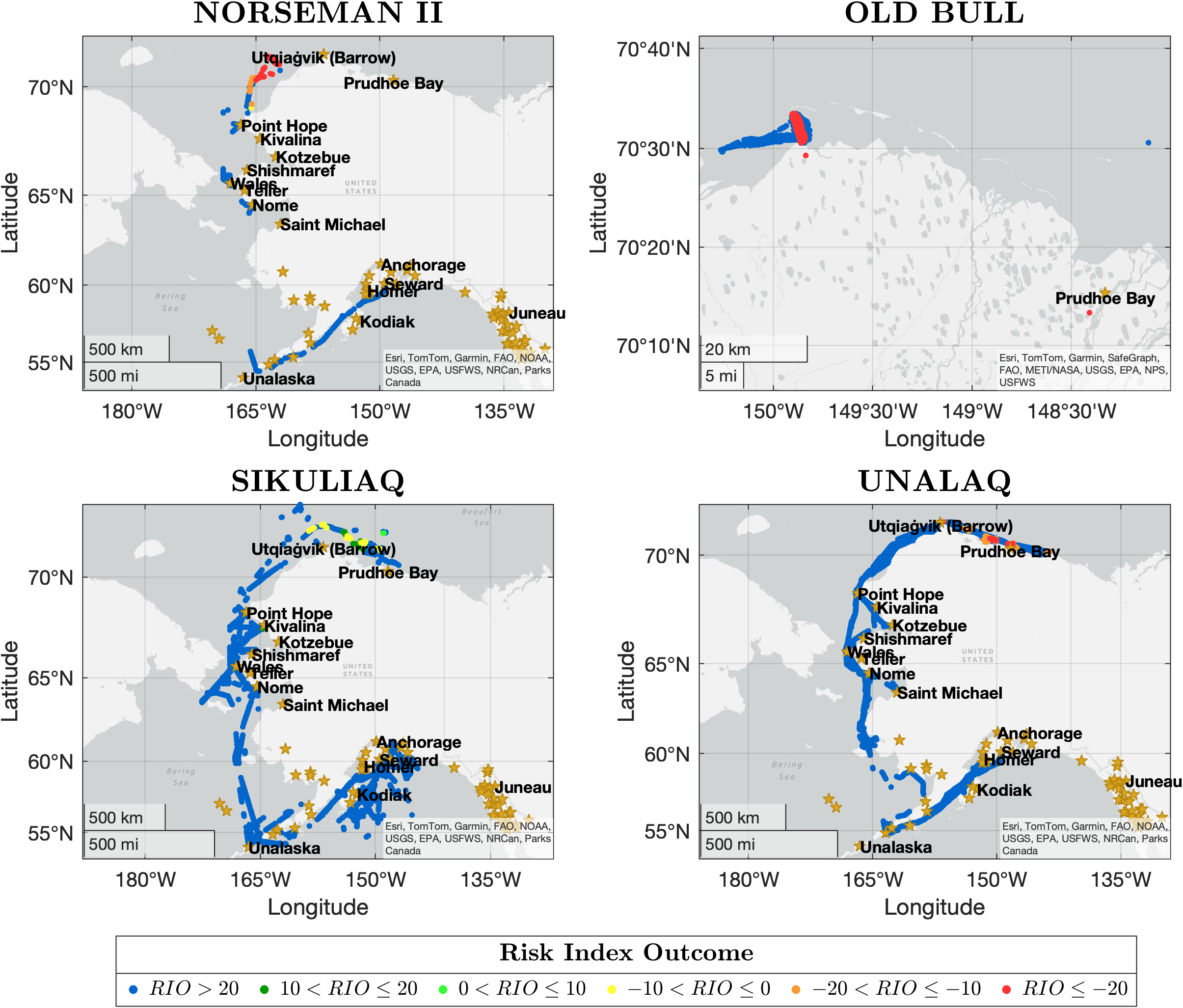}
\caption{Spatial distribution of Risk Index Outcomes (RIO) for the ships Norseman II (top left), Old Bull (top right), Sikuliaq (bottom left) and the Unalaq (bottom right). Negative RIO values indicate elevated operational risk. Gold stars represent public Alaskan port facilities.}
\label{fig:rio_map_2}
\end{figure}

Eight representative vessels—ranging from research ships and tugs to cargo vessels—were selected for the 2015–2020 period based on AIS coverage and proximity to ASIP-identified ice fields. This group includes two Polar Class ships (\textit{CG Healy}, \textit{Sikuliaq}) and six NIS vessels (\textit{Andy B}, \textit{BBC Oregon}, \textit{Emmett Foss}, \textit{Old Bull}, \textit{Norseman II}, and \textit{Unalaq}). Figures~\ref{fig:rio_map_1} and~\ref{fig:rio_map_2} show the spatial distribution of RIO values for these vessels, while Table~\ref{tab:polaris_summary} summarizes the corresponding POLARIS metrics.

Most vessels, particularly the tugs and the \textit{Unalaq}, operated frequently in areas where RIO values were negative, indicating elevated operational risk. Speed exceedances above recommended limits were also common, ranging from 40\% to 100\% of recorded AIS points for several vessels. These exceedances occurred most frequently in the Bering and Chukchi Seas during spring and autumn, when the marginal ice zone is most dynamic. Because the accuracy of RIO assessments depends directly on the spatial resolution and categorical definition of ice conditions, the choice of ice dataset is critical to quantifying operational risk.

Across all vessels combined, approximately 36\% of AIS observations recorded within ice-affected waters corresponded to negative RIO values, indicating operations under elevated-risk conditions. This widespread occurrence of sub-zero RIOs underscores that non–ice-strengthened vessels routinely operate in ice regimes exceeding POLARIS-defined safe thresholds. The concentration of these negative RIO points along recurring ice-edge zones in the Bering and Chukchi Seas further highlights the operational importance of accurate regional ice data. Unlike the Copernicus dataset, which provides continuous fractional ice concentrations unsuitable for direct integration into the categorical SIGRID framework, the ASIP charts supply the necessary standardized ice codes and classifications. This compatibility allows ASIP data to be seamlessly applied within the POLARIS RIO framework, producing a more realistic depiction of navigational risk exposure for Arctic vessels.

The primary limitation of this POLARIS analysis is that vessel selection was based on AIS–ASIP overlap from a single representative date (8/18/2018 as shown in Figure~\ref{fig:utm_overlay}) rather than continuous sampling across the full study period. As a result, the assessed vessels provide a cross-sectional snapshot of potential operational risks rather than a statistically complete record of Arctic shipping activity. In addition, POLARIS does not account for human or situational factors such as crew experience, vessel maneuverability, or real-time ice awareness. ASIP ice charts represent conditions at approximately 00:30~UTC, whereas vessel operations occur throughout the day; thus, temporal offsets between chart publication and actual navigation may contribute to apparent exceedances. Nevertheless, the prevalence of negative RIO values among non–ice-strengthened vessels highlights the importance of incorporating high-resolution, regionally validated ice data like ASIP into operational risk frameworks and route-planning tools. Future work should expand this framework to evaluate vessel operations over multi-year timescales, allowing for seasonally and interannually consistent assessment of POLARIS-based risk conditions.

\begin{table}[h!]
\caption{Vessel Characteristics}
\label{tab:vessel_info}
\centering
\begin{tabular}{llccccc}
\toprule
Name & MMSI & Length (m) & Beam (m) & Draft (m) & Type & Polar Class \\
\midrule
ANDY B & 367172150 & 18.35 & 5.7 & -- & Tug & NIS \\
BBC OREGON & 305462000 & 138.5 & 21.0 & 7.5 & Tanker & NIS \\
CG HEALY & 303902000 & 128.0 & 24.0 & 8.5 & Icebreaker & PC4 \\
EMMETT FOSS & 367576720 & 23.0 & 9.0 & 2.16 & Tug & NIS \\
NORSEMAN II & 367176270 & 33.0 & 8.8 & 4.0 & Research & NIS \\
OLD BULL & 367492440 & 11.0 & 5.0 & 1.68 & Tug & NIS \\
SIKULIAQ & 338417000 & 70.2 & 15.0 & 5.95 & Research & PC5 \\
UNALAQ & 338718000 & 45.0 & 15.2 & -- & Cargo & NIS \\
\bottomrule
\end{tabular}
\end{table}

\begin{table}[h!]
\caption{POLARIS Analysis Summary for Selected Vessels}
\label{tab:polaris_summary}
\centering
\resizebox{\textwidth}{!}{%
\begin{tabular}{lrrrrrllrrr}
\toprule
\textbf{Ship Name} & 
\textbf{Total AIS} & 
\textbf{AIS Pts} & 
\textbf{Neg.} & 
\textbf{AIS Pts} & 
\textbf{\% Speed} & 
\textbf{AIS Date} & 
\textbf{Days} & 
\textbf{Days Ex.} & 
\textbf{\% Speed} \\
& \textbf{Points} & 
\textbf{($\geq$1\%)} & 
\textbf{RIO} & 
\textbf{(SOG$\geq$limit)} & 
\textbf{Exceed.} & 
\textbf{Range} & 
\textbf{Neg. RIO} & 
\textbf{Speed Lim.} & 
\textbf{Exceed. (Days)} \\
\midrule
ANDY B & 105{,}278 & 34{,}511 & 9{,}701 & 256 & 3\% & 06/11/2017--10/14/2019 & 28 & 14 & 50\% \\
BBC OREGON & 56{,}027 & 297 & 8 & 8 & 100\% & 08/27/2017--12/12/2019 & 1 & 1 & 100\% \\
CG HEALY & 216{,}266 & 5{,}933 & 0 & 0 & --- & 01/01/2017--12/04/2019 & 0 & 0 & --- \\
EMMETT FOSS & 219{,}119 & 61{,}010 & 32{,}876 & 6{,}970 & 21\% & 07/03/2017--10/21/2019 & 56 & 49 & 88\% \\
NORSEMAN II & 8{,}673 & 767 & 720 & 436 & 61\% & 06/15/2016--07/14/2016 & 7 & 7 & 100\% \\
OLD BULL & 192{,}492 & 86{,}344 & 25{,}222 & 2{,}641 & 10\% & 07/07/2017--10/16/2019 & 36 & 28 & 78\% \\
SIKULIAQ & 346{,}881 & 1{,}450 & 164 & 65 & 40\% & 01/22/2017--12/30/2019 & 8 & 4 & 50\% \\
\bottomrule
\end{tabular}
}
\end{table}

\section{Conclusions}
\label{sec:conclusions}

This study compared satellite-derived Copernicus sea ice concentration fields with analyst-interpreted Alaska Sea Ice Program (ASIP) charts to evaluate spatial and temporal discrepancies in ice representation across Alaskan maritime regions. Daily ice concentrations were aligned within a common UTM grid, and residuals were computed to quantify differences at regional and seasonal scales. Across the 2010–2025 period, Copernicus consistently underestimated ice concentration relative to ASIP, particularly in nearshore and marginal ice zones where mixed pixels, and complex coastlines reduce passive-microwave accuracy.

Empirical Orthogonal Function (EOF) analysis revealed that both datasets capture the same dominant spatiotemporal modes of Arctic sea ice variability, confirming their shared representation of large-scale physical processes. The first EOF mode in both datasets primarily reflects the annual freeze–thaw cycle, while the second mode isolates non-seasonal variability associated with marginal ice zone dynamics. These findings demonstrate that differences between datasets are not due to physical inconsistency, but rather to the contrasting spatial resolutions and data structures used to represent the ice field.

Operational analysis using the IACS POLARIS framework showed that the choice of ice dataset has a direct influence on assessed navigational risk. When ASIP data were integrated with vessel Automatic Identification System (AIS) tracks, approximately 36\% of observations within ice-affected waters corresponded to negative Risk Index Outcomes (RIOs), indicating elevated operational risk. Because the Copernicus dataset provides continuous fractional concentrations unsuitable for categorical SIGRID input, only ASIP charts can currently be applied directly within the POLARIS framework. The ability of ASIP to capture fine-scale coastal ice variability therefore makes it indispensable for regional risk modeling and vessel-route planning. 

Overall, this work demonstrates that combining large-scale satellite products with regional expert datasets provides a more complete representation of Arctic sea ice conditions for both scientific and operational applications. Future work should extend this framework to other Arctic regions, incorporate multi-year AIS datasets for seasonally consistent risk assessment, and develop automated routines for merging regional ice charts with near-real-time satellite data to enhance navigational safety in dynamic ice environments. 

\section{Acknowledgments}

This material is based upon work supported by the National Science Foundation, Navigating the New Arctic Idea under Grant No. 2127095. The authors thank Mary-Beth Schreck, Sea Ice Program Leader at the NWS Alaska Sea Ice Program, for her assistance in accessing and interpreting the ASIP datasets.

\bibliographystyle{elsarticle-num} 
\bibliography{cas-refs-self}

\end{document}